\documentclass{aa}  

\usepackage[normalem]{ulem}
\usepackage{natbib}
\usepackage{graphicx,textpos}
\usepackage[dvipsnames]{xcolor}
\usepackage{longtable,ltcaption}
\usepackage{txfonts}
\usepackage{hyperref}
\usepackage{rotating}
\usepackage[utf8]{inputenc}
\usepackage{gensymb}
\DeclareUnicodeCharacter{00A0}{~}
\bibpunct{(}{)}{;}{a}{}{,}
\newcommand{\Msun}{M$_{\odot}$}

\newcommand{\M}[1]{\textcolor{black}{{#1}}}

\begin{document}

\title{Binary evolution along the Red Giant Branch with BINSTAR:\\
The barium star perspective}

\author{A. Escorza\inst{1,2} \and
L. Siess\inst{2} \and
H. Van Winckel\inst{1} \and
A. Jorissen\inst{2}
}

\offprints{A. Escorza, \\ \email{ana.escorza@kuleuven.be}}

\institute{
    Institute of Astronomy, KU Leuven, Celestijnenlaan 200D, B-3001 Leuven, Belgium \\
    \email ana.escorza@kuleuven.be \and
    Institut d'Astronomie et d'Astrophysique, Universit\'{e} Libre de Bruxelles (ULB), CP 226, B-1050 Bruxelles, Belgium 
    }

\date{\today}

\abstract
{Barium (Ba), CH and extrinsic or Tc-poor S-type stars are evolved low- and intermediate-mass stars that show enhancement of slow-neutron-capture-process elements on their surface, an indication of mass accretion from a former \M{asymptotic giant branch (AGB)} companion, which is now a white dwarf (WD). Ba and CH stars can be found \M{in} the main-sequence (MS), the sub-giant, and the giant phase, while extrinsic S-type stars populate the giant branches only. \M{As these polluted stars evolve, they might be involved in a second phase of interaction with their now white dwarf companion.}
In this paper, we consider systems composed of a main-sequence Ba star and a WD companion when the former evolves along the Red Giant Branch (RGB). We want to determine if the orbital properties of the known population of Ba, CH, and S giants can be inferred from the evolution of their suspected dwarf progenitors.
For this purpose, we use the BINSTAR binary evolution code and model MS+WD binary systems, considering different binary interaction mechanisms, such as a tidally-enhanced wind mass-loss and a reduced circularisation efficiency. To explore their impact on the second RGB ascent, we compare the modelled orbits with the observed period and eccentricity distributions of Ba and related giants.
We show that, independently of the considered mechanism, there is a strong period cut off below which core-He burning stars should not be found in binary systems with a WD companion. This limit is shorter for more massive RGB stars and for more metal-poor systems. However, we still find a few low-mass short-period giant systems that are \M{difficult to explain with our models} as well as two systems with very high eccentricities.\\}


\titlerunning{Binary evolution along the RGB}
\authorrunning{Escorza et al.}

\maketitle

\section{Introduction}\label{sec:intro}

Interactions between the two stellar components of a binary system are often unavoidable when one of them reaches giant dimensions. In this contribution, we focus on low- and intermediate-mass stars that were subject to mass transfer from a binary companion and are detected as being chemically peculiar, with surface enhancement in slow-neutron-capture-process (s-process) elements \citep{SmithLambert88,SmithLambert90}. These elements are an indication of mass accretion from a companion that was formerly on the \M{asymptotic giant branch (AGB)} and is now a dim white dwarf (WD), that in most cases, is not directly observable (see, however, \citealt{Bohm-Vitense00,Gray11} for direct detection of these WD companions). 

These families of peculiar stars include barium (Ba) stars, CH stars, and extrinsic or technetium-poor S stars.
Ba stars are G- or K-type giants \citep{BidelmanKeenan51} or F-, G- or K-type main-sequence stars (e.g. \citealt{North00}; \citealt{Gray11}) with the above-mentioned s-process enhancement. CH stars are their population II analogues and are on average more metal-poor. They are also found in the giant \citep{Keenan42}, subgiant \citep{Bond74}, and main-sequence \citep{Escorza17} phases. Extrinsic S stars are the cooler counterparts of barium giants, cool enough to show ZrO and TiO absorption bands in their spectra (e.g. \citealt{Jorissen88}; \citealt{Jorissen98}).

The orbits of these binary systems have been intensively scrutinized. Orbital parameters have been determined for giant Ba and CH stars and for extrinsic S stars by \cite{McClureWoodsworth90}, \cite{Udry98}, and \cite{Jorissen95, Jorissen98, Jorissen16, Jorissen19}, among others. Additionally, the orbital parameters of almost 30 main-sequence and subgiant Ba and CH stars are also known \citep{North00,North20,Escorza19}. In the eccentricity-period \M{($e - \log P$)} diagram, these families of polluted stars occupy a period range from about 100~d to periods longer than 10$^{4}$~d. Most of the short-period ($P\lesssim$~1000~d) systems are in circular orbits while no circular orbit is found among giant systems with $P > 1000$~d. These orbital elements are, of course, the post-interaction values, which certainly differ from the initial conditions. A tendency emerges from these observations that core-He burning Ba or CH stars have globally longer periods than their less evolved dwarf counterparts (see Figs. 15 and 16 of \citealt{Escorza17}), a hint that some further orbital evolution occurred long after the pollution with s-process elements.

On the theoretical side, investigations by \cite{Pols03}, \cite{BonacicMarinovic08}, \cite{Izzard10}, \cite{Dermine13}, \cite{Abate13,Abate18}, \cite{Saladino18}, and \cite{Saladino19}, among others, have tried to reproduce the orbital properties of binary systems formed after binary interaction with an AGB companion. 
All these simulations face the same problem: due to tidal forces, models predict cirulcular orbits below a certain period. This predicted threshold period is much longer than what is observed. 
A mechanism that can pump the eccentricity up is needed, and different processes have been proposed in the references mentioned above. For example, \cite{BonacicMarinovic08} used a phase-dependent mass loss to explain the orbital elements of Ba stars. Following \cite{Artymowicz91} and \cite{ArtymowiczLubow94}, the interactions between a binary system and its circumbinary (CB) disc have also been explored as a possibility to explain the eccentricities of post-AGB (e.g. \citealt{Dermine13,Oomen20}) and subdwarf B-type (e.g \citealt{Vos15}) stars, among other types of systems.

There is no doubt that the stage of mass transfer between the now polluted star and its former AGB companion shaped the orbits of these systems. However, Ba, CH, and extrinsic S giants are themselves evolved stars, and they could have undergone a second stage of binary interaction with their current WD companion while they ascended the Red Giant Branch (RGB). In this contribution, we investigate the effect of binary interaction on the orbital parameters of systems composed of a polluted main-sequence star with a WD companion. We want to determine if the population of known Ba, CH and extrinsic S giants can be inferred from the evolution of their suspected Ba and CH dwarf progenitors. This problem might also offer interesting clues to understand the interaction between the now Ba star and its former AGB companion. 

The observational constraints that motivate this study are described in Sect. \ref{sec:obs}. The modelling methodology and input physics of the models are described in Sect. \ref{sec:mod}. In Sect. \ref{sec:res+disc}, we describe and analyse the outcome of the different evolutionary models, and in Sect. \ref{ssec:res.assum}, we discuss the validity of our approximations. Finally, in Sect. \ref{sec:res+obs}, we compare our results with the observations introduced in Sect. \ref{sec:obs} and in Sect. \ref{sec:sum+concl}, we summarise our conclusions.

\section{Observational constraints}\label{sec:obs}

\begin{figure}
\centering
\includegraphics[width=0.49\textwidth]{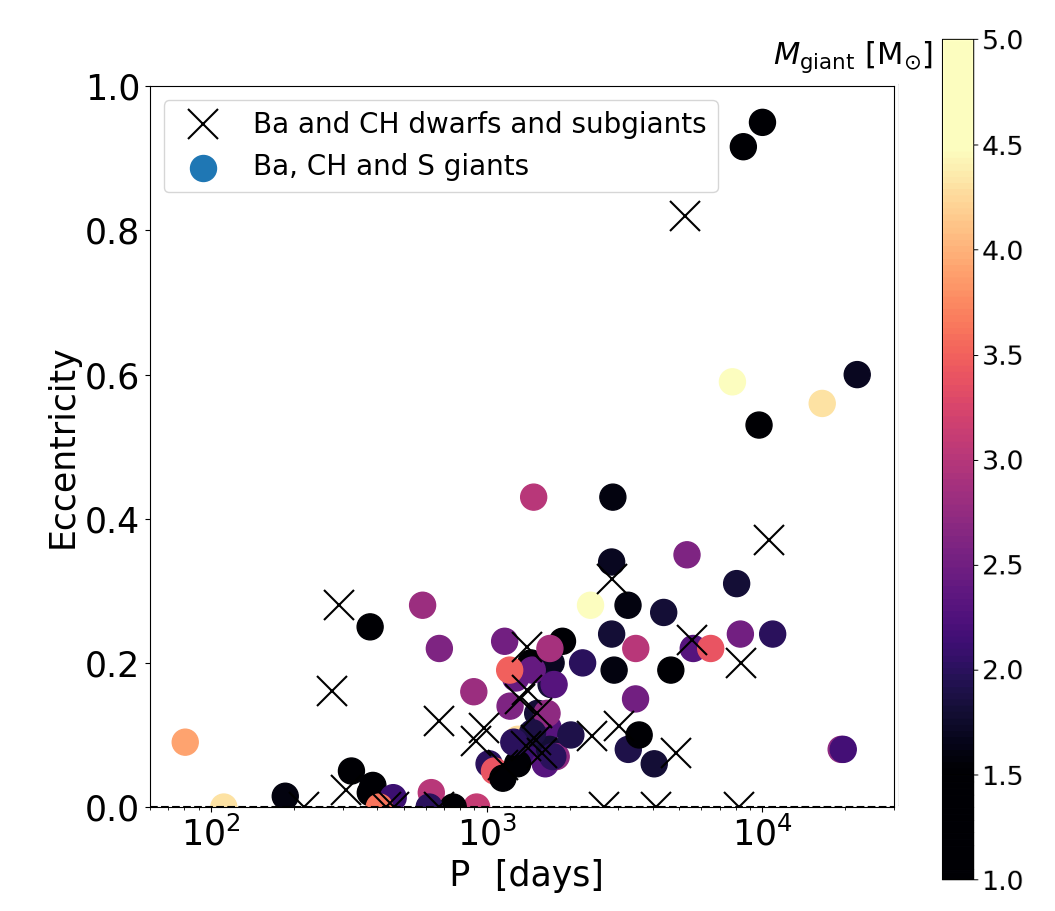}
\caption{\label{elogp_giants} Eccentricity-period diagram of a sample of Ba, CH and S giant systems with accurate masses. The colour of the symbol depends on the mass of the primary (Ba, CH or S) star, with the darkest symbols \M{associated with} the least massive stars and according to the colour scale on the right-hand side. The orbital elements of the progenitors, Ba and CH dwarfs and subgiants, are overplotted with black crosses.}
\end{figure}

Figure \ref{elogp_giants} shows the orbital parameters of the sample of polluted Ba, CH and S giants that will be used to compare with our models. The orbital parameters have been collected from  Table 4 of \cite{Jorissen16}, Tables A.1 to A.3 of \cite{VanderSwaelmen17}, and Table 4 of \cite{Jorissen19}. \M{The complete list of original references is included in Table \ref{orbits}.} In the figure, the systems are colour-coded as a function of the mass of the Ba, CH or S giant star and the darker the symbol, the lower the mass. The accuracy of the masses of these families of stars has improved a lot in the last few years after the two Gaia data releases \citep{Gaia-collaborationDR1,GaiaMission,Lindegren16,Lindegren18}. Only objects with accurate masses listed in Table 8 of \cite{Jorissen19}, in Table 4 of \cite{Shetye18}, or determined by us following their methodology, are included in the figure. \M{The masses of all the objects in Fig. \ref{elogp_giants} are also listed in Table \ref{orbits}.}

These systems will provide the basis to which we will confront different binary evolutionary models, computed with the BINSTAR code \citep{Siess13,Davis13}. \M{Note, however, that while we have orbital information for all the known Ba giants with strong anomalies (i.e., those classified as Ba3, Ba4, or Ba5 in the scale by \citealt{Warner1965}), the other samples are incomplete. Some stars classified as mild Ba, CH or Tc-poor S stars still have tentative orbits \citep{Jorissen19}.} The crosses in Fig. \ref{elogp_giants} indicate the orbital elements of the Ba and CH dwarfs and subgiants from Table 1 of \cite{Escorza19} and Table 1 of \cite{North20}.

\section{Modeling methodology}\label{sec:mod}
For our computations, we use the BINSTAR binary evolution code (see \citealt{Siess13} and \citealt{Davis13} for a detailed description of the code). BINSTAR solves the internal structure of the two components of a binary system and its orbital evolution (i.e. change of eccentricity and separation) at the same time. 

The modelled systems consist of a main-sequence star and a cool WD companion, and they are evolved until the end of the core-He burning phase of the primary star. The term \textit{primary} refers here to the star that starts the simulations on the main-sequence phase, because this is the most massive and usually most luminous star in the modelled systems. In order to speed up the calculations and avoid convergence problems, the models only follow the evolution of the stellar structure of the primary star. The WD is allowed to gain and lose mass, but it is treated as a point mass. We discuss the validity of this assumption in Sect. \ref{ssec:res.assum}. The coming subsections give more details about the specific stellar and binary physics considered in the models and about the grid set-up.

\subsection{Input stellar and binary physics}\label{ssec:phys}

The stellar input physics available in BINSTAR is inherited from the 1D stellar evolution code STAREVOL, and it is described in \cite{Siess2006}. \M{Convection in these models is based on the mixing length theory with $\alpha_{\rm MLT} = 1.75$ as calibrated for solar models. Our previous study \citep{Escorza19} suggested that some core overshooting was necessary to explain the location of Ba dwarfs in the Hertzsprung-Russell (HR) diagram (see Fig. 8 of \citealt{Escorza19}). Therefore, we included overshooting on top of the convective core following the exponential decay expression of \cite{Herwig97} with $f_{\rm{over}} = 0.01$.
We did not consider any additional mixing mechanism, although they are available in the code, because the evolution of the surface chemical abundances of the modelled stars is out of the scope of this work.}

\M{The surface boundary condition of the models is a grey atmosphere, and the mass loss is modelled according to \cite{SchroderCuntz07} as:}

\begin{equation}\label{eq:SchroderCuntz07}
\dot{M} = \eta \frac{L_{\ast} R_{\ast}}{M_{\ast}} 
\;
\left( \frac{T_{\rm eff, \ast}}{4000 
} \right)^{3.5} 
\;
\left( 1 + \frac{g_{\odot}}{4300 
\;
g_{\ast}} \right),
\end{equation}\

\noindent \M{where $L_{\ast}$,  $R_{\ast}$, and $M_{\ast}$ are the stellar luminosity, radius, and mass in solar units, $T_{\rm eff, \ast}$ is the stellar effective temperature (expressed in K), and $g_{\ast}$ and $g_{\odot}$ are the stellar and solar surface gravity, respectively. The value $\eta = 8 \times 10^{-14}$~\Msun\ yr$^{-1}$ was empirically estimated by \cite{SchroderCuntz05} to reproduce the mass loss of globular cluster RGB stars.}

Radiative opacities are taken from \cite{IglesiasRogers96} above 8000\,K and from \cite{Ferguson05} at lower temperatures. \M{We used the standard nuclear network included in BINSTAR, which includes 182 reactions coupling 55 species from H to Cl.} The chemical composition is scaled solar according to the \cite{Asplund09} mixture. The impact of metallicity on the results is discussed in Sect. \ref{ssec:res.assum}. 

For the binary modeling, the two stars are assumed to be rigid rotators and they can lose mass via stellar winds, which carries away the specific orbital angular momentum of the star.
The binary systems consist of a primary Ba star that at certain stages of its evolution will transfer mass to its WD companion. We use the subscript "Ba" for the donor star and the subscript "WD" for the secondary and gainer star. These stars have masses $M_{\rm Ba}$ and $M_{\rm WD}$, radii $R_{\rm Ba}$ and $R_{\rm WD}$, and angular velocities $\Omega_{\rm Ba}$ and $\Omega_{\rm WD}$. The total angular momentum of the system is the sum of the stellar angular momentum of the two stars ($J_{\rm Ba}$ and $J_{\rm WD}$) and the orbital angular momentum, $J_{\rm orb}$:
\begin{equation}\label{J}
J_{\rm \Sigma} = J_{\rm Ba} + J_{\rm WD} + J_{\rm orb},
\end{equation}\

\noindent and these quantities may change as a result of mass transfer, mass loss, and tidal interaction.

The evolution of the binary separation is given by 

\begin{equation}\label{adot}
\frac{\dot{a}}{a} = 2 \frac{\dot{J}_{\rm orb}}{J_{\rm orb}} - 2 \left( \frac{\dot{M}_{\rm Ba}}{M_{\rm Ba}} + \frac{\dot{M}_{\rm WD}}{M_{\rm WD}} \right) + \frac{\dot{M}_{\rm Ba}+\dot{M}_{\rm WD}}{M_{\rm Ba}+M_{\rm WD}} + \frac{2e\dot{e}}{1-e^2} .
\end{equation}\

The resolution of this equation requires prescriptions for the mass-loss and mass-transfer rates, the torque, $\dot{J}_{\rm orb}$, and rate of change of the eccentricity, $\dot{e}$. $\dot{J}_{\rm orb}$ is normally calculated by imposing the conservation of the total angular momentum (Eq. \ref{J}):
\begin{equation}\label{Jorbdot}
\dot{J}_{\rm orb} = \dot{J}_{\rm \Sigma} - \dot{J}_{\rm Ba} - \dot{J}_{\rm WD},
\end{equation}\

\noindent where $\dot{J}_{\rm \Sigma}$ is the rate at which angular momentum is lost from the system, by wind in our specific case. In the Jeans mode \citep{Jeans24, Jeans25}this is given by:
\begin{equation}\label{Jsigmadot}
\dot{J}_{\rm \Sigma} = \sum\limits_{i} \dot{M}^{\rm wind}_{i} a^2_i \omega = \left( \frac{\dot{M}^{\rm wind}_{\rm Ba}}{q} + q \dot{M}^{\rm wind}_{\rm WD} \right) j_{\rm orb},
\end{equation}\

\noindent where $\omega$ is the orbital angular velocity, $a_i$ is the distance of star $i$ to the center of mass of the system, $q = M_{\rm Ba}/M_{\rm WD}$ is the mass ratio, and $j_{\rm orb} = J_{\rm orb}/(M_{\rm Ba}+M_{\rm WD})$ is the specific orbital angular momentum. The rates of change of the stellar angular momenta ($\dot{J}_{\rm Ba}$ and $\dot{J}_{\rm WD}$) include contributions from mass loss, mass accretion and tides \citep[for details see][]{Siess13}. In our models $\dot{M}^{\rm wind}_{\rm WD} = 0$.\\

Finally, the total rate of change of the eccentricity is the sum of the following eccentricity-changing factors:
\begin{equation}\label{edot}
\dot{e} = \dot{e}_{\rm wind} + \dot{e}_{\rm tide},
\end{equation}\

\noindent where $\dot{e}_{\rm wind}$ depends on the orbital phase via the true anomaly, $\nu$ \citep{Soker00,BonacicMarinovic08}:
 
\begin{equation}\label{edotwind}
\dot{e}_{\rm wind} (\nu) = \frac{\vert \dot{M}^{\rm wind}_{\rm Ba}(\nu) + \dot{M}^{\rm wind}_{\rm WD}(\nu) \vert}{M_{\rm Ba}+M_{\rm WD}} \left( e + \cos \nu\right).
\end{equation}\

\M{This relation expresses that a variable mass-loss rate along the orbit implies that the angular-momentum loss becomes phase-dependent as well. This asymmetry in the angular-momentum loss induces some variation of the eccentricity. If the mass-loss rate were constant, the integral of Eq. \ref{edotwind} over an orbital period is zero, so the eccentricity then remains constant. In order to follow the evolution of the system over millions of orbits when the mass-loss or mass-transfer rates depend on the orbital phase, BINSTAR uses averaged quantities. The treatment of phase-dependent rates in BINSTAR is described in \cite{Siess14}.}

\M{Finally, the tidal term, $\dot{e}_{\rm tide}$, has been derived by \cite{Hut81} using the timescales estimated by \cite{Zahn77,Zahn89}:}

\begin{equation}\label{edottidepaper}
\frac{\dot{e}_{\rm tide}}{e} = - \frac{\left( 1 - e^2 \right)^{-13/2}}{\tau_{\rm circ}}
\left[ \frac{18}{7} f_3(e^2) - \frac{11}{7} \left( 1 - e^2 \right)^{3/2} f_4(e^2) \frac{\Omega_{\rm spin}}{\omega} \right], 
\end{equation}

\noindent \M{where $f_{\rm 3}$ and $f_{\rm 4}$ are polynomials in $e^2$, and the circularisation timescale, $\tau_{\rm circ}$, is expressed as:}

\begin{equation}\label{taucircpaper}
\frac{1}{\tau_{\rm circ}} = 21 \frac{\lambda_{\rm 10}}{t_{\rm f}} \tilde{q} (1 + \tilde{q}) \left(\frac{R}{a}\right)^8,
\end{equation}\

\noindent \M{where $\lambda_{\rm 10}$ is a parameter related to the harmonic used in the expansion of the tidal potential (see \citealt{Zahn89} for its derivation), $\tilde{q} = M_{\rm i} / M_{\rm 3-i}$ is the ratio of the mass of the companion to the mass of the considered star, and the friction time is $t_{\rm f} = (MR^2/L)^{1/3}$ where $M$, $R$, and $L$ are the mass, the radius, and the luminosity of the considered star.}

\subsection{Tidally-enhanced wind mass loss}\label{ssec:CRAP}
\cite{ToutEggleton88} suggested that the gravitational pull of a companion could enhance the wind mass loss. They proposed a prescription where mass loss is driven by tidal interaction and depends on the orbital separation through the ratio between the radius of the mass-losing star and its Roche lobe radius, $R/R_{\rm L}$. In an eccentric orbit, $R_{\rm L}$ changes with the orbital phase because the separation is not constant\footnote{Note that the usage of a phase-dependent $R_{\rm L}$ deviates from the original conditions of the Roche Model which assumed circular orbits.}, and the enhanced mass-loss prescription becomes phase-dependent as well:

\begin{equation}\label{MdotCRAP}
\dot{M}_{\rm wind} (\nu) = \dot{M}_{\rm ref} \times \left\lbrace 1 + B_{\rm wind} \times \min \left[ \left(  \frac{R}{R_{\rm L} (\nu)} \right)^{6}, \frac{1}{2^{6}} \right]  \right\rbrace,  
\end{equation}\

\noindent where $\dot{M}_{\rm ref}$ is the reference mass-loss rate (from \citealt{SchroderCuntz07} in our case), $R$ is the stellar radius, $R_{\rm L}(\nu)$ is the Roche-lobe radius that depends on the true anomaly (via the instantaneous separation), and $B_{\rm wind}$ is a free parameter.

$B_{\rm wind}$ is not well constrained and is generally adjusted to reproduce the properties of specific systems. For example, \cite{ToutEggleton88} estimated $B_{\rm wind} = 10^4$ to account for the properties of the RS CVn system Z Her, and \cite{Siess14} estimated $B_{\rm wind} = 3.6 \times 10^4$ to model the long-period eccentric system IP Eri.
In our models, we test different values of $B_{\rm wind}$ and analyse its impact on the orbital evolution along the RGB.

\subsection{Reduced circularisation efficiency}\label{ssec:redtide}

\cite{Nie17} suggested that a reduced circularisation efficiency is needed to explain the existence of ellipsoidal red giants in eccentric binary systems in the Large Magellanic Cloud. They studied 81 systems that had distorted, high-luminosity RGB primaries and found that about 20\% of them had higher eccentricities than predicted by tidal circularisation theory, while the rest were in circular orbits. They suggested that these systems could be accounted for if tidal circularisation rates were about two orders of magnitude smaller than the standard values.

Since we also find eccentric systems involving primary stars that are on the RGB, we explore this scenario as well. To account for this reduced efficiency, the value of $\dot{e}$ associated with tides in Eq. \ref{edot} is multiplied by a factor $F_{\rm tide} = 0.01$, \M{as suggested by \cite{Nie17}}, and keeping the contribution from the wind unaltered:
\begin{equation}\label{edotF}
\dot{e} = \dot{e}_{\rm wind} + F_{\rm tide} \times \dot{e}_{\rm tide,Ba} +  F_{\rm tide} \times \dot{e}_{\rm tide,WD}\ .
\end{equation}

\subsection{Model grid set-up}

The initial period and eccentricity ranges used in the models were chosen so they could lead to the formation of Ba-giant systems like those observed. The explored parameter space is then the following:

\begin{itemize}
    \item Four primary star masses: 1.5~M$_{\odot}$, 2.0~M$_{\odot}$, 2.5~M$_{\odot}$, and 3.0~M$_{\odot}$,
    
    \item six initial orbital periods: 100\,d, 300\,d, 600\,d, 1000\,d, 2000\,d, and 3000\,d,
    
    \item \M{four initial eccentricities: 0.1, 0.2, 0.4, and 0.6.}
\end{itemize}

\M{Even though Ba stars have masses up to 5~M$_{\odot}$ (Fig. \ref{elogp_giants}), we focus on the low-mass part of the distribution ($M \leq 3$~M$_{\odot}$). As will be shown in Sect. \ref{sec:res+disc}, main-sequence stars with masses $M \geq 2.5$~M$_{\odot}$ do not interact with their WD companion unless they are in very short orbits. We set the lowest initial mass in our grid to 1.5~M$_{\odot}$, because very few Ba giants have been observed with masses as low as 1~M$_{\odot}$. Furthermore, the near solar composition of the Ba giants suggests that they have formed not too long ago. Indeed, if the clump was populated with 1~M$_{\odot}$ stars, these stars would have formed 10 billion years ago and at the time, their metallicity was likely lower than the currently observed value.} With this set-up, we are assuming that the orbital properties of main-sequence Ba and CH stars are similar for all masses. This is an extrapolation since only Ba dwarfs with $M \lesssim$~1.6~M$_{\odot}$ have been detected observationally. This is quite different from the Ba giants, where a broader mass distribution centred around 2~M$_{\odot}$ is observed. The lack of more massive Ba dwarfs is attributed to observational biases \citep{Escorza19}. \M{Concerning the orbital elements, we restrict our grid to periods shorter than 3000~d, because wider systems will only interact if they have very high eccentricities.}

Additionally, stars in our comparison sample cover a non-negligible metallicity range. Ba stars have a narrow metallicity distribution that peaks at about [Fe/H]~$=-0.14 \pm 0.2$ \citep{deCastro16,Escorza17} and the metallicities of extrinsic S stars fall in this range as well \citep{Shetye18}, but CH stars are more metal-poor. Our adopted solar metallicity does not differ significantly from the average metallicity of Ba stars and these constitute the bulk of our comparison sample. However, we should keep this aspect in mind when comparing individual objects. The effect of metallicity is discussed in Sect. \ref{ssec:res.assum}.

The initial mass of the WD companion is set to 0.59~M$_{\odot}$ in all the models. This value corresponds to the value at which the WD mass distribution of the SDSS \citep{Kleinman13} and the Gaia DR2 \citep{Tremblay19} catalogues peak. \M{Note that these are mass distributions of single WDs, but \cite{Jorissen19} showed that the mass distribution of WD companions of Ba giants peaks at a similar value although it is slightly broader than that of single WDs and has a tail extending toward larger masses. The impact of using a more massive WD will be discussed in Sect. \ref{ssec:res.assum}.}

All models are evolved up to the end of the core He-burning phase of the primary star. In some systems, the primary star leaves the RGB prematurely due to the strong interaction with its WD companion and never ignites helium. These systems will thus not lead to the He-clump Ba and CH giants that we observe but are briefly discussed in Sect. \ref{ssec:res.std}.

\begin{figure*}
\centering
\includegraphics[width=\textwidth]{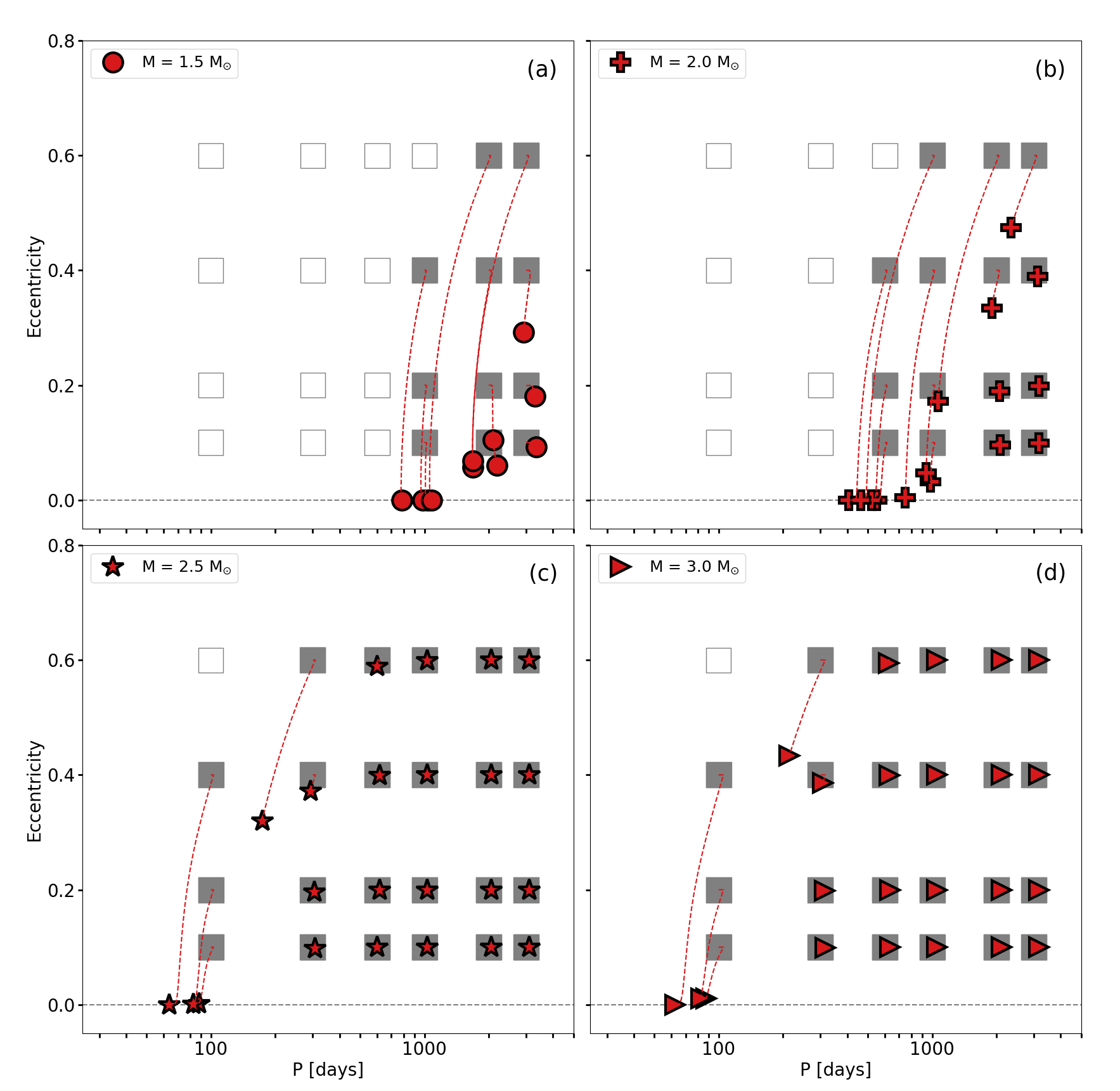}
\caption{\label{nocrap_ALL} Evolution in the eccentricity-period diagram of systems that follow standard binary evolution. The dashed-red line indicates the evolution of the orbit between the initial orbital parameters (grey rectangles) and the orbital parameters of the system when core-He burning starts (red symbols). Open grey rectangles represent systems where the primary does not ignite helium in the core because it leaves the RGB prematurely. Each panel corresponds to a different primary initial mass.}
\end{figure*}

\section{Results of the simulations}\label{sec:res+disc}

\subsection{Standard binary evolution}\label{ssec:res.std}

Figure \ref{nocrap_ALL} shows the evolution in the eccentricity-period diagram of the models for which we used standard binary evolution. Grey rectangles represent the initial orbital parameters of each model and the red symbols represent the system's properties at the onset of core-He burning. We chose to analyse our models at the onset of core-He burning because afterward, the orbits mainly widen due to mass loss. This choice allows us to explore the lower limit of the orbital periods allowed for RGB and core-He burning stars due to tidal interaction along the RGB.

The dashed lines follow the evolution of the orbital parameters for those systems that reach the core-He burning stage. The different panels correspond to the different initial primary masses. In panels (c) and (d), we see that the stars with the highest initial masses only interact with their WD companion if the initial period is equal or shorter than 300\,days. \M{The primary-star radius} remains small with respect to the Roche-lobe radius of these wide orbits. On the other hand, stars with initial masses equal to 1.5 and 2~M$_{\odot}$ suffer from much stronger interaction and orbital circularisation. This is because the least massive stars reach much larger radii at the RGB tip, up to more than 175~R$_{\odot}$ for a star with initial mass of 1.5~M$_{\odot}$. Hence, tidal circularisation is much stronger. Between systems with the same initial period, tidal circularisation is stronger in those with a higher initial eccentricity. More eccentric systems reach smaller separations at periastron passage, decreasing the value of $R_{\rm L}$ and increasing the effect of the tides.

\begin{figure*}
\includegraphics[width=\textwidth]{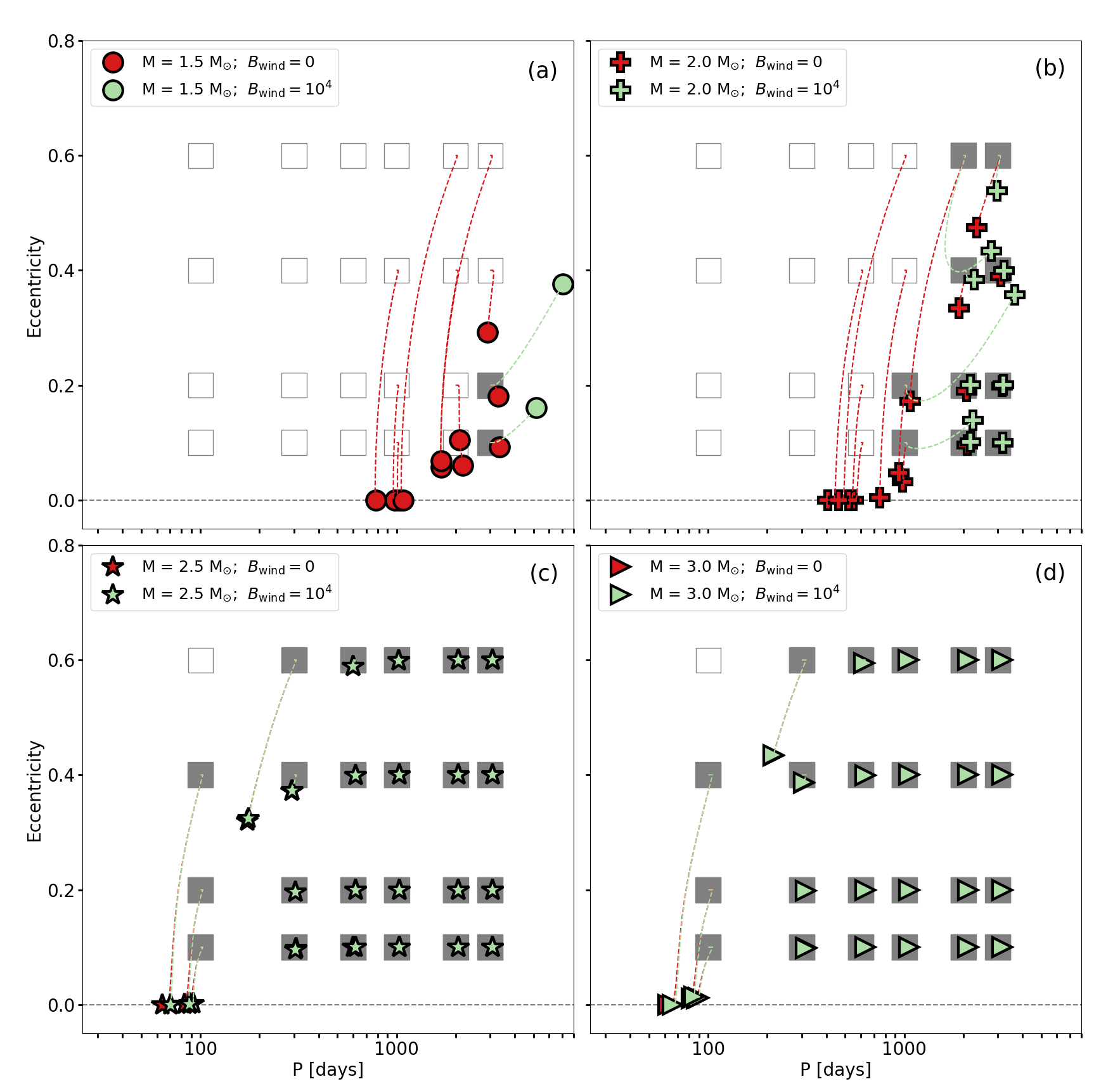}
\caption{\label{no+104_ALL} Same as Fig. \ref{nocrap_ALL} but including systems with a tidally-enhanced wind mass loss with $B_{\rm wind}\,=\,10^4$ in green. Open grey rectangles indicate the systems, among those with an enhanced wind, where the primary does not ignite its He core and leaves the RGB prematurely.}
\vspace{0.3cm}
\end{figure*}

We also see that for a considerable fraction of low-mass stars and for more massive stars in very short and eccentric systems, binary interaction is strong enough to make them leave the RGB prematurely, thus avoiding core-He burning. This is the fate of all the systems in Fig. \ref{nocrap_ALL} displayed as open grey rectangles, indicating the initial orbital parameters. They do not have an evolutionary dashed line and a final orbit symbol associated with them. These systems undergo mass transfer via RLOF onto the WD companion with mass-accretion rates of the order of $10^{-6} - 10^{-5}$~M$_{\odot}$~yr$^{-1}$. As mentioned in Sect. \ref{sec:mod}, we do not follow the structural changes of the WDs, but given the magnitude of these rates, the accreted matter is expected to pile up onto the compact object, expand and lead to the formation of a common envelope (CE) \citep[e.g.][]{Wolf13}. In order to assess the fate of these systems after the CE phase, we determined the final orbital separation of each system after the removal of the envelope of the RGB star \M{(e.g. \citealt{Paczynski76,LivioSoker88,Han95,Hurley02}).} The binding energy of the donor's envelope $E_{\rm bind}$ is computed using the stellar structure information from our models and is given by:

\begin{equation}\label{Ebind}
E_{\rm bind} = - \int^{M_{\rm Ba}}_{M_{\rm Ba}^{\rm core}} \left(\frac{Gm}{r} - U_{\rm int}(m) \right) {\rm d}m,
\end{equation}

\noindent where we integrate over the mass coordinate $m$ from the base of the envelope ($M_{\rm Ba}^{\rm core}$) to the surface of the Ba star \M{at the onset of the common-envelope phase ($M_{\rm Ba}$).} In the above expression, $G$ is the gravitational constant, $r$ is the radial coordinate, and $U_{\rm int}(m)$ is the specific internal energy.

Assuming that the variation of orbital energy during the CE phase is enough to expel the loose envelope of the giant, we can estimate the final separation $a^{\rm f}$ from the following relation:

\begin{equation}\label{CE?}
E_{\rm bind} = E^{\rm f}_{\rm orb} - E^{\rm i}_{\rm orb}  = 
- \frac{G}{2}
\left( \frac{M^{\rm core}_{\rm Ba}M^{\rm f}_{\rm WD}}{a^{\rm f}}
- \frac{M^{\rm i}_{\rm Ba}M^{\rm i}_{\rm WD}}{a^{\rm i}} \right),
\end{equation}\

\noindent where $M^{\rm core}_{\rm Ba}$ and $M^{\rm i}_{\rm Ba}$ are the final and initial Ba star masses, $M^{\rm f}_{\rm WD}$ and $M^{\rm i}_{\rm WD}$ are the final and initial WD masses, and $a^{\rm i}$ is the separation at the onset of RLOF. 

\begin{figure*}
\centering
\includegraphics[width=\textwidth]{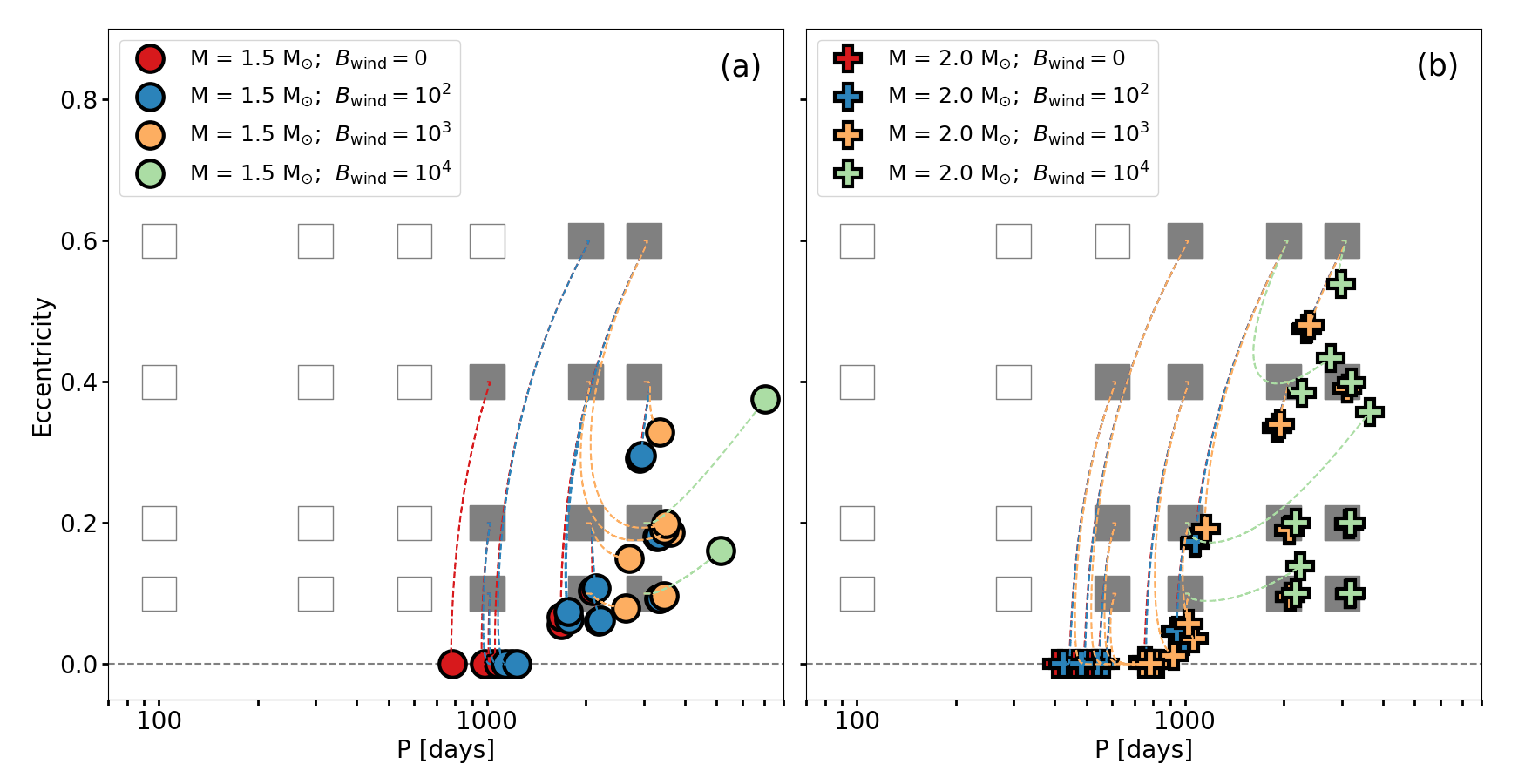}
\caption{\label{CRAPS_1.5_2.0} Same as Figs. \ref{nocrap_ALL} and \ref{no+104_ALL} but only for the systems systems with initial mass $M=$\,1.5 and 2.0\,M$_{\odot}$. The figures show different values of $B_{\rm wind}$: 0 in red, $10^2$ in blue, $10^3$ in yellow, and $10^4$ in green. Open grey rectangles indicate the systems that do not start core-He burning for any value of $B_{\rm wind}$, including $B_{\rm wind} = 0$.}
\end{figure*}

Using this relation we find that after the envelope is ejected, the core of the Ba star is still filling its Roche lobe, so we conclude that the Ba and CH dwarfs that suffer RLOF while ascending the RGB will likely merge and produce a sub-Chandrasekhar massive WD.

\M{The details of the common-envelope evolution require detailed modelling that is beyond the scope of this publication; however, we can estimate that, if the core of the Ba star merges with its WD companion, the product will contribute to populate the high-mass tail of the WD mass distribution found by  \cite{Hollands18} and \cite{Kilic18} using Gaia DR2 data and that \cite{Kilic18} attributed to mergers. Alternatively, studies have proposed that the accretion of He-rich material onto a WD can lead to He-burning detonation, which could trigger core ignition and subsequent Type Ia supernova (e.g. \citealt{Shen14,Shen15,Townsley19}). This could also be a possible post-CE outcome for the merger of the He-core of a low-mass Ba star with its WD companion.}

\subsection{Tidally-enhanced wind mass loss}\label{ssec:res.crap}

Figure \ref{no+104_ALL} shows the orbital evolution of models with a tidally-enhanced wind mass loss with $B_{\rm wind}\,=\,10^4$ (green) and compares it with the orbital evolution of the standard binary models (red). The first result is that an enhanced wind only affects systems with low-mass primaries ($M \lesssim 2.0~$M$_{\odot}$), for the same reasons as those discussed in Sect. \ref{ssec:res.std}. The overlapping red and green symbols in panels (c) and (d) show that the orbital evolution of the systems with primary stars of masses 2.5~M$_{\odot}$ and 3.0~M$_{\odot}$ is unaffected by the enhanced mass loss. This is due to the fact that a more massive donor star expands less on the RGB than their lower-mass counterparts, which makes the filling factor $R/R_{\rm L}$ in Eq. \ref{MdotCRAP} smaller.

Because of the enhanced mass-loss rate, in systems with primary stars of 1.5~M$_{\odot}$ and 2.0~M$_{\odot}$, the eccentricity-pumping term $\dot{e}_{\rm wind}$ exceeds $\dot{e}_{\rm tide}$ and when the primary star ignites helium, these systems still have a significant eccentricity. However, the stronger wind also makes many more systems leave the RGB before core-He burning starts. For example, among all the binaries with a 1.5~M$_{\odot}$ primary (panel (a) of Fig. \ref{no+104_ALL}), only the system with an initial period equal to 3000~d and an initial eccentricity equal to 0.2 reaches core-He ignition.

Contrary to the standard case ($B_{\rm wind}\,=\,0$, Sect. \ref{ssec:res.std}), the systems that do not reach the core-He burning phase neither suffer from RLOF nor merge. In this case, the RGB star loses most of its envelope via the enhanced stellar wind and this prevents the primary star from overflowing its Roche lobe. These stars leave the RGB prematurely to form a double degenerate system (He-WD + CO-WD).

\begin{figure*}
\includegraphics[width=\textwidth]{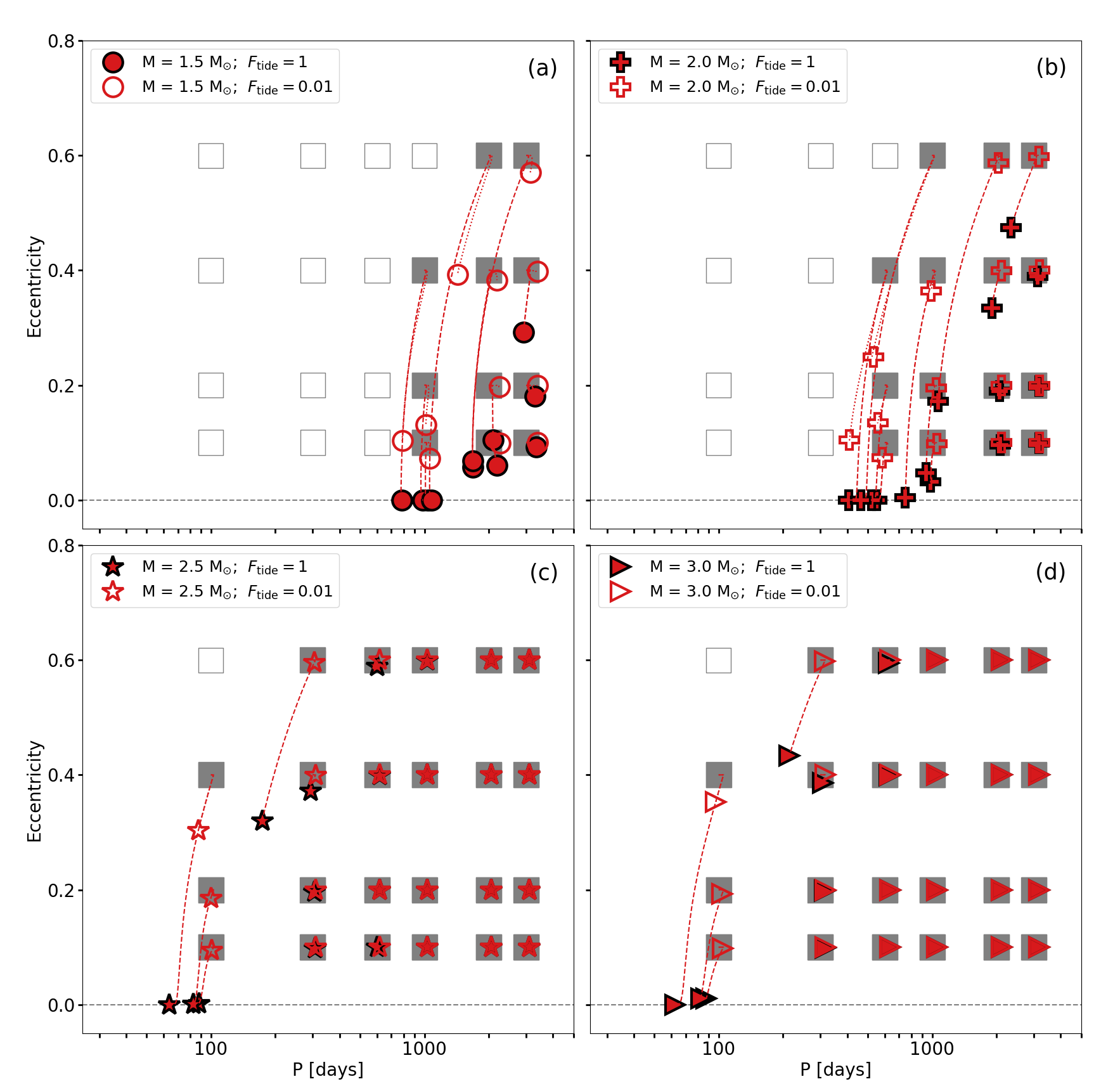}
\caption{\label{no+tide_ALL} Same as Fig. \ref{nocrap_ALL} but including systems with a reduced circularisation efficiency $F_{\rm tide} = 0.01$ (open red symbols). Open grey rectangles mark systems where the primary does not ignite He and leaves the RGB prematurely.}
\end{figure*}

To analyse further the impact of $B_{\rm wind}$, we tested two additional values ($10^2$ and $10^3$) on the low-mass systems, where the effects of the tidally-enhanced wind are significant. Figure \ref{CRAPS_1.5_2.0} shows, as expected, an intermediate behaviour of the orbital parameters. With a lower $B_{\rm wind}$, more systems survive the envelope peel-off during the ascent of the RGB. For example, none of the models with initial mass 1.5~M$_{\odot}$ and periods of 1000 or 2000~d reach core-He burning if $B_{\rm wind} = 10^4$, but some of them do when $B_{\rm wind}$ is lower. Additionally, the final eccentricity of a system with given initial orbital parameters increases with increasing $B_{\rm wind}$. Only models with $B_{\rm wind} \leq 10^2$ produce low-mass (with $M_{\rm init}=1.5~$M$_{\odot}$) core-He burning stars in circular orbits, but this is an artifact of our initial conditions, since we did not consider circular systems at the start of the simulation. The behaviour of the systems with initial mass $M=2.0~$M$_{\odot}$ is similar, but more systems survive the envelope peel-off during the RGB phase and shorter periods are allowed at the core-He burning stage. \M{Again this is because the radius of a 2.0~M$_{\odot}$ star of solar metallicity at the tip of the RGB is about 125~R$_{\odot}$, significantly smaller than that of a 1.5~M$_{\odot}$ star, which is about 175~R$_{\odot}$ (see Fig. \ref{radii}, discussed later in the text and \citealt{Escorza17err}).}

\subsection{Reduced circularisation efficiency}\label{ssec:res.tides}

Figure \ref{no+tide_ALL} shows the orbital evolution of the models with a reduced circularisation efficiency (open symbols) together with the standard binary models of Fig. \ref{nocrap_ALL} (filled symbols). All the systems with massive primaries, $M=$~2.5~M$_{\odot}$ in panel (c) and $M=$3.0~M$_{\odot}$ in panel (d), keep their eccentricities until core-He ignition, even at very short initial periods. For binaries with lower initial primary masses (panels a and b), most systems that used to circularize in the standard case retain some eccentricity until the core-He burning phase. This reduction of the tidal efficiency cannot, however, prevent RLOF and the same initial conditions that led to mergers in the case with $F_{\rm tide}=1$ also lead to mergers with $F_{\rm tide}=0.01$. 

\subsection{The combined effect}

\begin{figure*}
\centering
\includegraphics[width=\textwidth]{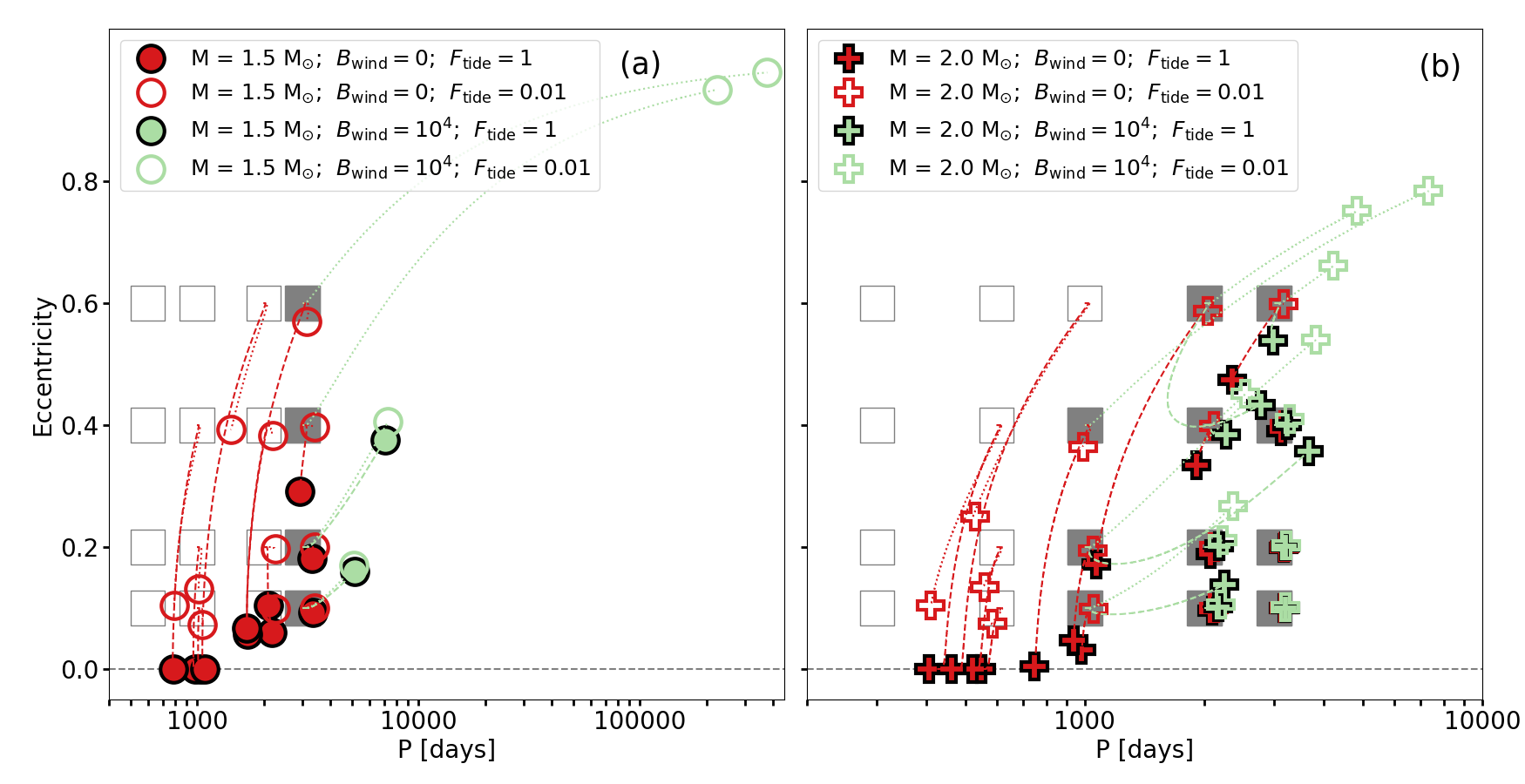}
\caption{\label{tides+crap_15_20} Same as Fig. \ref{no+tide_ALL} but only for systems with initial mass $M=$\,1.5 and 2.0\,M$_{\odot}$. The figures show models with two different values of $B_{\rm wind}$ ($B_{\rm wind}=$~0 in red and $B_{\rm wind}=10^4$ in green) and two different circularisation efficiencies ($F_{\rm tide} = 1$ with filled symbols and $F_{\rm tide} = 0.01$ with open symbols). Grey rectangles indicate initial orbital parameters and the filled rectangles mark those systems that reached core-He burning when a tidally-enhanced mass loss and a reduced tidal efficiency are combined ($B_{\rm wind}=10^4$ and $F_{\rm tide} = 0.01$).}
\end{figure*}

We also analysed the combined effect of a tidally-enhanced mass loss and a reduced tidal efficiency. Fig. \ref{tides+crap_15_20} compares the standard binary models ($B_{\rm wind}=$~0 and $F_{\rm tide}=1$, filled red symbols), the models with a tidally-enhanced wind ($B_{\rm wind}=10^4$, filled green symbols), the models with a reduced circularisation efficiency ($F_{\rm tide} = 0.01$, open red symbols), and a new set of models with $B_{\rm wind}=10^4$ and $F_{\rm tide} = 0.01$ (open green symbols).

When the processes are combined, more systems with low-mass primaries reach the He-core burning phase than when the tidally-enhanced wind alone is considered (there are two more empty green symbols than filled green symbols in the left panel and one more in the right one). When the strength of the tidal interaction is reduced, the orbits remain wide for a longer time and mass lost due to the tidally-enhanced wind also gets reduced. This makes more systems finish the RGB with a high enough mass to ignite the helium core. However, the orbital parameters of these systems at He-core ignition are very extreme and these effects empty the $e-\log P$ diagram with systems in the period range $\lesssim 2000$~d. The two sets of systems, with initial masses $M =$~1.5 and 2.0~M$_{\odot}$, end up in very high eccentricities and long periods.

\begin{figure}
\centering
\includegraphics[width=0.49\textwidth]{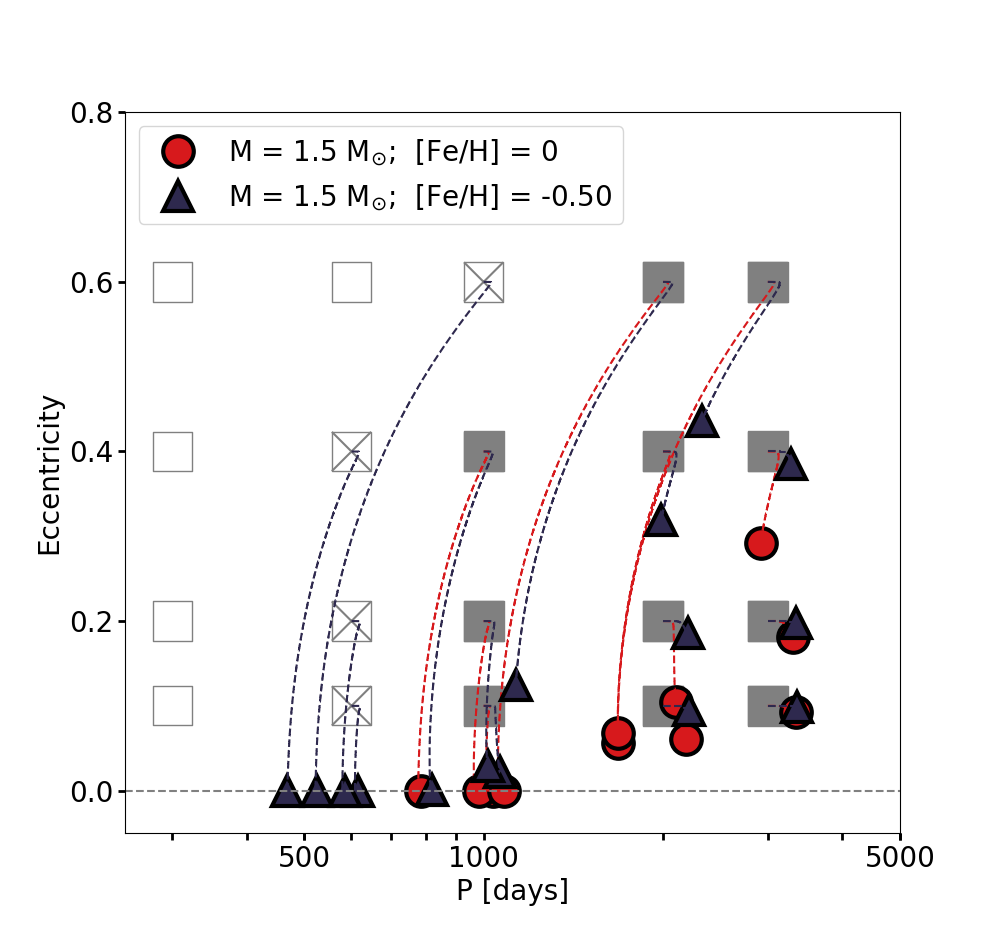}
\caption{\label{mettest} Orbital evolution of systems with initial mass $M=1.5$~M$_{\odot}$ at [Fe/H]~$=0$ (red circles) and [Fe/H]~$=-0.50$ (blue triangles). Filled rectangles mark systems that reach core-He burning independently of the metallicity. Models marked with a crossed rectangle ignite the He core only at sub-solar metallicity.}
\end{figure}

\subsection{Validity of the assumptions}\label{ssec:res.assum}
 In this section, we analyse the impact of the most important assumptions and approximations made in this analysis.

\begin{itemize}

\item[\textbf{(i)}] \textbf{Metallicity.} All our models have solar metallicity, but the metallicity distribution of Ba stars peaks at [Fe/H]~$=-0.14 \pm 0.2$ (Fig. 4 of \citealt{Escorza17}). Additionally, there are some CH giants in the comparison sample, and they are more metal-poor.
In order to investigate the effect of metallicity on the final orbital elements, we compare, in Fig. \ref{mettest}, the orbital evolution of systems with initial mass $M=1.5$~M$_{\odot}$ and solar metallicity (red circles) with those having [Fe/H]~$=-0.5$ (blue triangles). As before, grey rectangles indicate initial orbital parameters, and they are filled when a system reaches the core-He burning phase with both metallicities. We see that more binaries survive the ascent of the RGB and reach the core-He burning stage when the metallicity is lower. \M{This is because metal-poor stars are more compact and reach smaller radii at the RGB tip, so shorter periods are required for RLOF to occur. Figure~\ref{radii} shows the radii at the tip of the RGB for stars of different initial masses and at two different metallicities. This plot illustrates why metal-poor stars interact at shorter orbital periods, but also why the modelled systems with primaries more massive than 2.0~M$_{\odot}$ barely interact.} The effect of metallicity on binary interaction along the RGB could explain the accumulation of metal-poor giants with 300\,d $ \lesssim P \lesssim$ 600\,d observed by \cite{Jorissen04, Jorissen19EWASS} in the $e- \log P$ diagram. We did not investigate the regime of more extreme metallicities.\\

\begin{figure}
\includegraphics[width=0.49\textwidth]{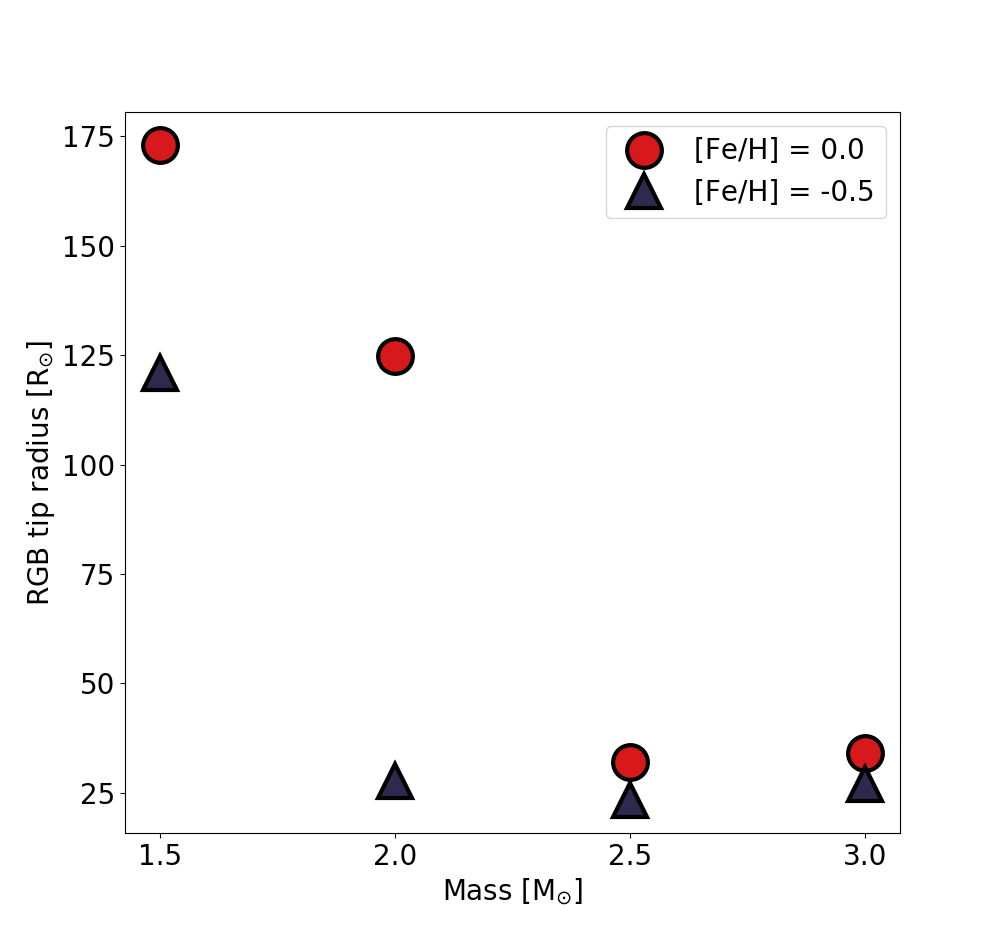}
\caption{\label{radii} \M{Radii at the tip of the RGB for stars with different initial masses and metallicities (see also \citealt{Escorza17err}).}}
\end{figure}

\item[\textbf{(ii)}] \textbf{WD mass.} The mass of the WD in all models was chosen to be 0.59~M$_{\odot}$ based on SDSS and Gaia WD mass distributions. Considering a more massive WD companion would impact the orbital evolution of the systems mainly via tidal interaction. According to equations 9 and 10 of \cite{Hut81}, $\dot{a}_{\rm tide}$ and $\dot{e}_{\rm tide}$ depend on $- \tilde{q} (1+ \tilde{q})$ with $\tilde{q} = M_{\rm WD} / M_{\rm Ba}$. Hence, a more massive WD companion, i.e. a higher value of $q$, would lead to faster circularisation rates and stronger shrinkage of the orbit. \M{However, this effect is not significant considering the evolutionary time scales. A 1.5~M$_{\odot}$ star in the same orbit but with a 1~M$_{\odot}$ WD companion instead would circularise only a factor of two faster and reach the clump in a slightly smaller orbit.\\}

\item[\textbf{(iii)}] \textbf{WD internal structure.} We did not take into account the evolution of the structure of the WD companion. This will mainly impact systems that suffer from RLOF because the WD accretes mass at very high rates. However, these systems are likely to merge \citep{Wolf13}.\\

\item[\textbf{(iv)}] \textbf{Circumbinary (CB) discs.}  CB discs can potentially pump the orbital eccentricity up \citep{ArtymowiczLubow94}. They have been invoked to explain the large eccentricities often observed among post-interaction binaries (e.g. \citealt{Dermine13} used a CB disc to explain the eccentricities of post-AGB binaries, and \citealt{Vos15}, to explain sub-dwarf B-type stars in long and eccentric orbits). The presence of CB discs in post-AGB or post-RGB binaries is inferred from the excess in the mid- and far-infrared seen in their spectral energy distributions, which is attributed to thermal re-emission by dust (\citealt{deRuyter06}; \citealt{Kamath16}; \citealt{vanWinckel17}; \citealt{Oomen18}). \M{In \cite{Escorza17} we analysed the spectral energy distributions of about 400 stars flagged as Ba or CH giants in different catalogues, and none of them showed an IR excess, including those still ascending the RGB. This means that if a disc forms at all around these systems, its dissipation time scale is probably much shorter than the evolutionary timescale of the RGB and core-He burning phases. Hence, since there is no evidence for the presence of CB discs around Ba giants, we did not consider them in our RGB evolutionary models, despite the fact that such discs may play a crucial role during the former mass-transfer episode leading to the formation of the Ba dwarf.}

\end{itemize}

\section{Comparison with observations}\label{sec:res+obs}

\begin{figure*}
\centering
\includegraphics[width=0.49\textwidth]{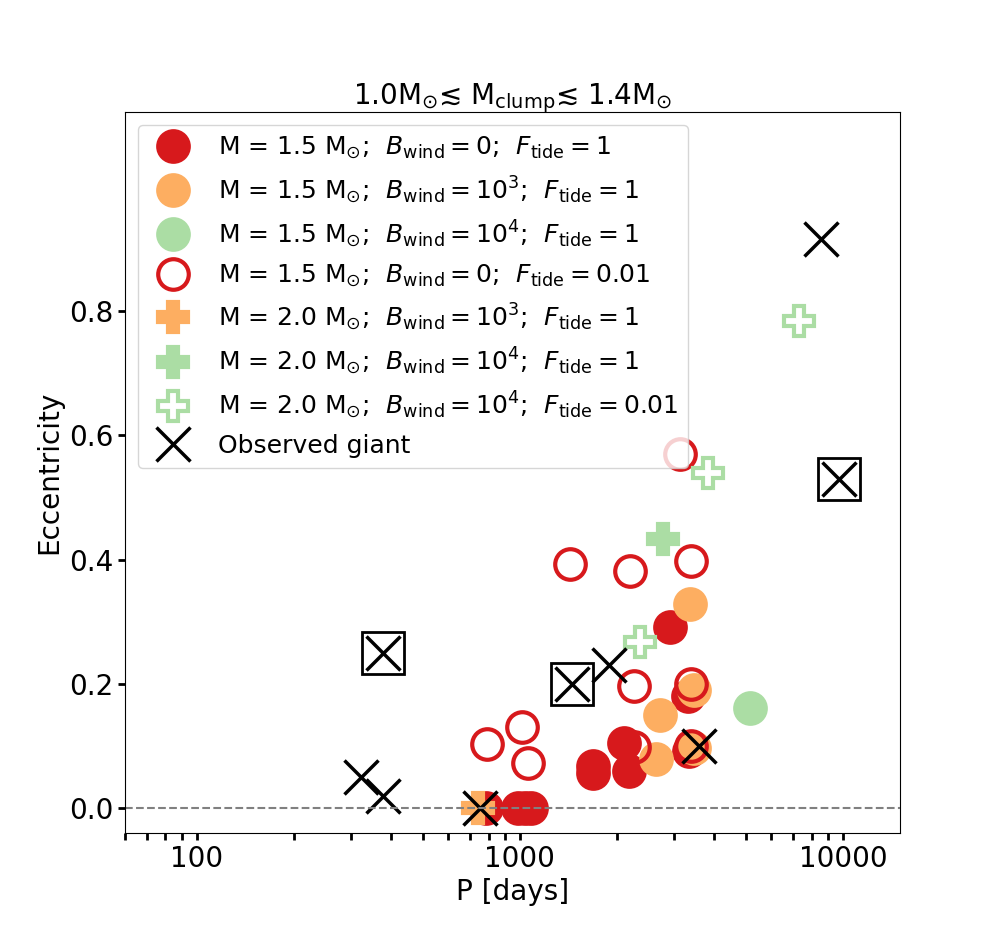}
\includegraphics[width=0.49\textwidth]{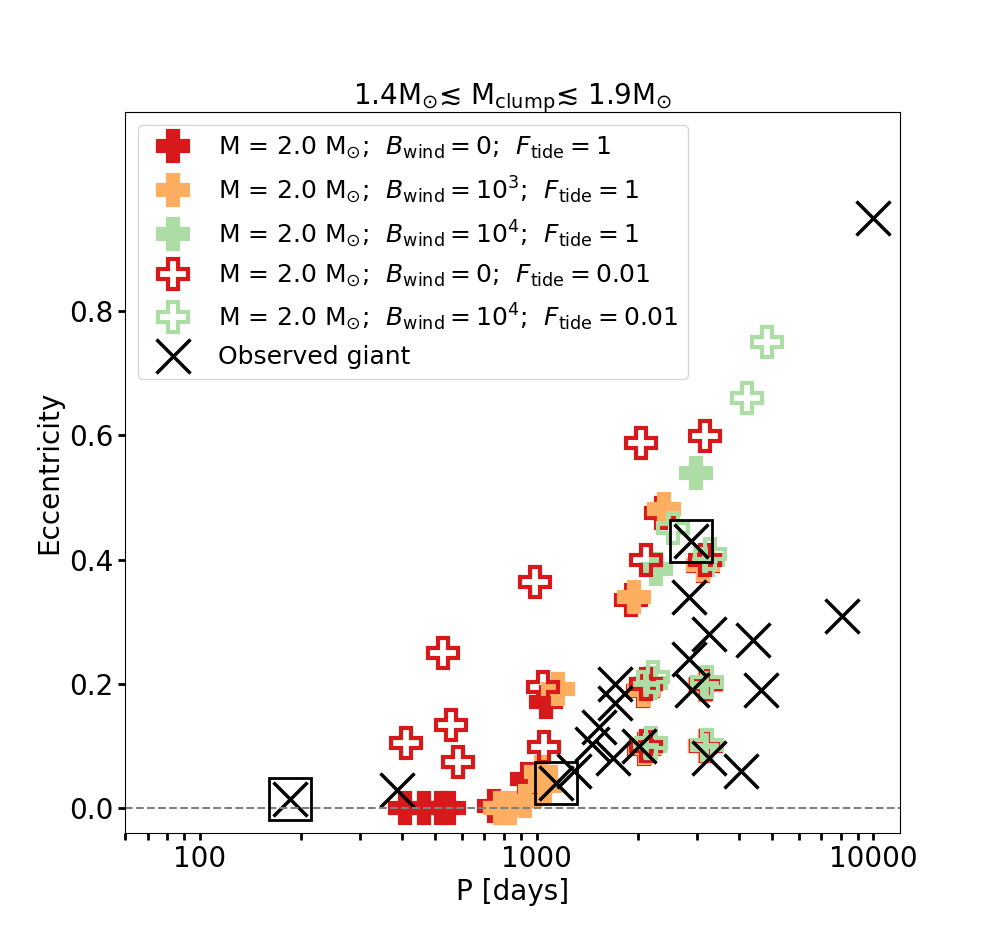}
\caption{\label{withdata} Eccentricity-period diagram of Ba giants \citep{Jorissen16,Jorissen19, VanderSwaelmen17} compared with the final (i.e., located in the red clump) orbital parameters of different sets of models analysed in Sect. \ref{sec:res+disc}. As before, different colours represent different values of $B_{\rm wind}$ and open symbols mark models with $F_{\rm tide} = 0.01$. Stars marked with \M{squares} are clearly located along the RGB or early AGB and should therefore not be compared with our predictions valid for red-clump stars only.}
\end{figure*}

\begin{figure}
\centering
\includegraphics[width=0.49\textwidth]{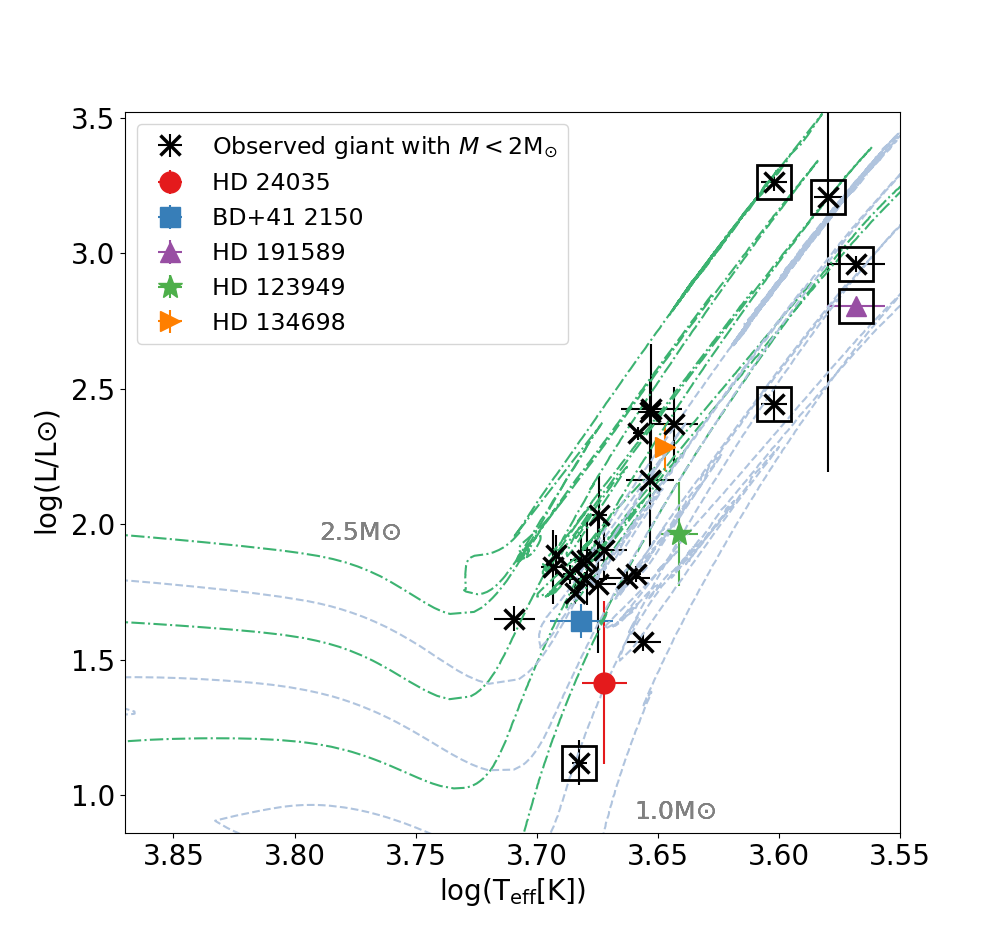}
\caption{\label{hrd} Hertzsprung-Russell diagram of low-mass Ba giants. The Ba, CH or S giants discussed in the text are plotted with different symbols. Stars that are clearly on the RGB or AGB and are not core-He burning stars are surrounded by a \M{square}. STAREVOL evolutionary tracks for 1.0, 1.5, 2.0 and 2.5~M$_{\odot}$ and with [Fe/H]~$\sim$~-0.5 (dash-dotted green tracks) and [Fe/H]~$\sim$~0 (dashed blue line) are included as reference.}
\end{figure}

Figure \ref{withdata} compares the $e- \log P$ diagram resulting from the different simulations with observations of low-mass Ba, CH, and S giants. The comparison has been divided in two mass bins based on the mass of the simulated stars when they start burning He in their cores, $M_{\rm clump}$. The cut-off masses of 1.4 and 1.9~M$_{\odot}$ are the masses that the 1.5 and 2~M$_{\odot}$ primary stars have, respectively, when core-He burning starts in the models. As discussed before, during the core-He burning phase, tidal interaction is not as strong as it is along the RGB ($R/R_{\rm L}$ decreases when the star ignites He) and stars lose mass via stellar winds, so the orbits widen. Hence, adopting $M_{\rm clump}$ and selecting the models at the beginning of the core-He burning phase gives a lower limit on the expected orbital periods after the RGB evolution. Additionally, since Fig. \ref{withdata} shows simulated orbits when the primary star ignites helium, Fig. \ref{hrd} has been used to identify stars that are not core-He burning stars and exclude them from the comparison. Stars that, taking the observational errors into account, are clearly RGB (or early AGB) stars are marked in Figs. \ref{withdata} and \ref{hrd} with \M{big squares}. The evolutionary tracks included in Fig. \ref{hrd} cover the mass range from 1.0 to 2.5~M$_{\odot}$ and were computed with the STAREVOL evolution code \citep{Siess2006} with two different metallicities ([Fe/H]~=~0 with dashed blue lines and [Fe/H]~=~-0.5 with dash-dotted green lines).

Among the models with different $B_{\rm wind}$ values, models with $B_{\rm wind}=100$ are not displayed because the final orbits are very similar to the final orbits of models with $B_{\rm wind}=0$ (see Fig. \ref{CRAPS_1.5_2.0}). Additionally, as discussed in Sect. \ref{sec:res+disc}, systems with $M_{\rm init}$~=~2.5 and 3.0~M$_{\odot}$ keep their main-sequence orbital parameters unless they are in very short orbits, so these are not included in the comparison either. Finally, models in which the primary star reaches $M<$~0.8~M$_{\odot}$ at some point of its evolution are also excluded because we do not observe Ba or CH giants of such low masses \citep{Jorissen19}. This is the case of only a few models where the star loses most of its outer layers due to the tidally-enhanced wind.

\setlength{\tabcolsep}{4pt}
\renewcommand{\arraystretch}{1.4}
\begin{table*}
\caption{\label{table}Period $P$, eccentricity $e$, metallicity [Fe/H], mass $M$, and corresponding references, for five Ba, CH, or S giants with outlying orbital properties according to Fig. \ref{withdata}. Their location on the HR diagram is shown in Fig. \ref{hrd}.}
\begin{center}
\begin{small}
\begin{tabular}{ccccccccc}
\hline
\rule[0mm]{0mm}{4mm}
ID 1 & ID 2 & Type & $P$ [days] & $e$ & Orbit ref. & [Fe/H] & $M$ [M$_{\odot}$] & $M$ ref.\\
\hline
HD\,24035 & HIP\,17402 & Ba & 377.8 $\pm$ 0.3 & 0.02 $\pm$ 0.01 & \cite{Udry98} & -0.23 & 1.3 $\pm$ 0.3 & \cite{Jorissen19}\\ 
BD+41$^\circ$2150 & HIP\,53832 & CH & 322.84 $\pm$ 0.08 & 0.055 $\pm$ 0.001 & \cite{Jorissen16} & -0.8 & 1.27 $\pm$ 0.2 & This work$^{(1)}$ \\
HD\,123949 & HIP\,69290 & Ba & 8523 $\pm$ 8 & 0.9162 $\pm$ 0.0003 & \cite{Jorissen19} & -0.23 & 1.3 $\pm$ 0.3 & \cite{Jorissen19}\\
HD\,134698 & BD-09$^\circ$4082 & Ba & $\sim$10005 & $\sim$0.95 & \cite{Jorissen19} & -0.57 & 1.5 $\pm$ 0.2 & \cite{Jorissen19}\\
HD\,191589 & HIP 99312 & S & 377.3 $\pm$ 0.1 & 0.250 $\pm$ 0.003 & \cite{Udry98} & -0.3 & 1.0 & \cite{Shetye18}\\
\hline
\end{tabular}
\end{small}
\end{center}
$^{(1)}$ Following the method used by \cite{Jorissen19} and \cite{Escorza19}.
\end{table*}

Figure \ref{withdata} shows that there is a sharp period cut off below which stars with a WD companion should not be observed after their RGB evolution. At solar metallicity, this period is about 750 days for stars below 1.5~M$_{\odot}$ and about 400 days for stars between 1.5 and 2.0~M$_{\odot}$. This cut off depends, however, on metallicity and it goes down to 450 days for $M<$~1.5~M$_{\odot}$ and [Fe/H]~$=-0.5$ (see Fig. \ref{mettest}). \M{On the long-period range, we observe that our models preserve the low-eccentricity gap observed in Ba giants with $P > 1000$~d (see \citealt{Jorissen04}). Note, however, that our input parameters did not include $e=0$.} Finally, as discussed in Sect. \ref{ssec:res.crap}, the figure shows that models with a tidally enhanced mass-loss do not produce orbits with short enough periods, leaving many low-mass giant systems unexplained. It also shows that a reduced circularisation efficiency produces orbits that are more eccentric than the observed ones.

\M{In Fig.~\ref{withdata}, two objects with $M\,<\,1.4\,\rm{M}_{\rm \odot}$ (left panel) and orbital periods below 400\,days cannot be reproduced by our binary evolution models, because they have shorter periods than the cut off set by RLOF. These are HD\,24035 and BD+41$^\circ$2150. Taking into account their orbital periods (Table \ref{table}), their location in the HR diagram (Fig. \ref{hrd}), and the results of our simulations, we can now conclude that they are likely early RGB stars. Their orbital parameters suggest that they will probably go through RLOF and merge before reaching the core-He burning phase.} Additionally, their periods are close to one year, which means that the Gaia DR2 parallaxes computed with the single-star solution could be unreliable (\citealt{PJ2000}, \citealt{Jorissen19EWASS}). Since these parallaxes have been a crucial ingredient to locate the stars in Fig. \ref{hrd} and to determine their masses, we need to be cautious with our conclusions until the Gaia DR3 catalogue is published. 

There are two additional stars, one in each mass bin, with long periods and very high eccentricities that the models cannot reproduce. They are HD\,123949 and HD\,134698, two Ba giants with orbits recently constrained by \cite{Jorissen19}. We computed a few additional standard binary models at their masses and metallicities, and with an initial period of 9000\,days and an eccentricity of 0.9 the system will reach the clump with $P_{\rm clump} \sim$ 7000~d and $e_{\rm clump} \sim$ 0.7, so these stars might also be ascending the RGB and will lose some eccentricity after they complete their RGB evolution. The period, eccentricity, metallicity, and mass of the mentioned stars are available in Table \ref{table} together with the corresponding references.

Finally, there is an additional outlier in the $e-\log P$ diagram which is worth mentioning even though is not a clump star. HD\,191589 is a S star of 1.0~M$_{\odot}$ \citep{Shetye18} that has a very short period (377~d) and a very high eccentricity (0.25) for its location in the HR diagram. \M{None of our binary evolution paths passes through its location in the $e-\log P$ diagram. This system has been reported to be intriguing before, because its mass function is 0.394~$\pm$~0.005~M$_{\odot}$ \citep{Udry98}, which is barely compatible with a WD companion. Only a very high inclination, of almost 90$^{\circ}$, is possible for a system with a 1~M$_{\odot}$ primary and such a mass function, and this would imply a very massive WD of almost 1.3~M$_{\odot}$. Finding such a massive WD as a companion of a low-mass polluted star would imply a quite extreme initial mass ratio. Additionally, it would challenge the idea that low-mass AGB stars are responsible for the pollution of Ba stars (e.g. \citealt{Lugaro16,Cseh18,GorielySiess18,Karinkuzhi18}). HD\,191589 is not the only extrinsic S star challenging our binary evolution models. For example, $o^1$ Ori is not included in our comparison sample because its orbital elements are only tentative \citep{Jorissen19}. However, this thermally-pulsing AGB star has a mass of 1.5~M$_{\odot}$ \citep{Shetye20}, a tentative period of 575~d, and a tentative eccentricity of 0.22. Hence, if further studies confirm this orbit, $o^1$ Ori would also lie in a region of the $e- \log P$ diagram that our models cannot populate.}

\M{We have not yet commented on the evolution of the surface abundances of Ba stars along the RGB, but during the first dredge-up, the star becomes convective and the surface material mixes in and dilutes. In our 1.5 and 2~M$_{\odot}$ models, at its deepest extent during the first dredge-up, the envelope reaches a mass coordinate of 0.27 and 0.30~M$_{\odot}$, respectively, implying that $\sim$~80-85\% of the star becomes well mixed. This process could decrease the surface s-process enhancement between the main-sequence and the core-He burning phases, perhaps enough to remove some barium star from the sample of giants. However, additional mixing mechanisms can affect the surface abundances of the polluted stars on the main sequence after the mass transfer episode. For example, the effect of thermohaline mixing has been investigated in the framework of Carbon-Enhanced Metal-Poor stars (CEMP stars; e.g. \citealt{Stancliffe07, Stancliffe08, Stancliffe09, Aoki08}) and Ba stars (e.g. \citealt{Proffitt89,Husti09}). \cite{Stancliffe08} found that thermohaline mixing can mix the accreted matter over 16 to 88\% of the mass of a 0.8~M$_{\odot}$ CEMP star with this percentage depending on the amount of accreted material and the mass of the polluter. Assuming that these estimates can be applied to our 1.5 and 2~M$_{\odot}$ stars, in the case of very efficient mixing (over 88\% of the mass), the surface abundance of s-elements will not be affected by dilution during the first dredge-up. On the other hand, if we take their lower limit (accreted material mixed over 16\% of the mass), the surface abundance of s-elements could still be diluted by a factor of $\sim 4$ (i.e. $-1.6$ dex) during the first dredge up, likely removing some of Ba dwarfs from the giant sample and reclassifying some others from strong to mild in the scale by \citealt{Warner1965}. Finally, rotationally-induced mixing (e.g. \citealt{Denissenkov00}) or atomic diffusion (e.g. \citealt{Matrozis16,Matrozis17}) might also contribute to the chemical evolution of polluted stars, but their efficiency remains to be fully assessed.}

\section{Summary and conclusions}\label{sec:sum+concl}

In this paper, we investigated whether a second stage of binary interaction has any effect on the eccentricity-period diagram of Ba and other polluted giants. With the BINSTAR binary evolution code, we evolved systems consisting of a main-sequence star and a cool WD companion and study binary interaction along the RGB of the former. The initial orbital elements were chosen to cover those of observed Ba and CH dwarfs. We explored three different cases: 

\begin{itemize}
\item In the standard binary evolution, low-mass ($M \lesssim 1.4$~M$_{\odot}$) core-He burning giants only appear in systems with periods greater than 750~d at solar metallicity. The low-mass ($M \lesssim 1.4$~M$_{\odot}$) Ba stars that start their RGB ascent in short-periods systems ($P \lesssim 800$~d for solar metallicity) will merge with their WD companions.\\

\item A tidally-enhanced wind mass-loss with $B_{\rm wind} = 10^4$ as proposed by \cite{ToutEggleton88} makes most stars with main-sequence masses $M_{\rm init} \lesssim 2$~M$_{\odot}$ leave the RGB prematurely. The systems  that survive the RGB ascent have longer periods ($P>2000$\,days) leaving a considerable fraction of observed intermediate-period Ba giants unexplained.\\

\item Finally, a reduced circularisation efficiency as proposed by \cite{Nie17} produces systems with higher orbital eccentricities when the primary star reaches the clump, but the period cut off below which RLOF occurs is the same as when using the standard tidal theory. Additionally, in systems with Ba giants with $M \gtrsim 1.5$~M$_{\odot}$, the final eccentricities are significantly higher than the observed ones.
\end{itemize}

We can conclude that while binary interaction along the RGB is negligible for Ba giants with $M \gtrsim 2.5$~M$_{\odot}$, low-mass Ba stars might go, depending on their main-sequence orbital period, through a second phase of interaction with their WD companions while they evolve along the RGB. We report as well that the limiting period below which RLOF is expected to lead to a merger strongly depends on the mass of the evolving Ba star and its metallicity. This is not a surprise, since both more massive and more metal-poor stars reach smaller radii on the RGB-tip, which reduces the tidal interaction.

\M{Our RGB binary evolution models can explain most of the observed orbits of core-He burning Ba giants using the known population of Ba dwarfs as initial conditions and taking into account the evolution of the Ba star itself along the RGB. Additional mechanisms not explored in this publication, such as interactions with a third star in the system, might need to be invoked to reproduce a few individual systems that remain unexplained.}

\begin{acknowledgements}
This research has been funded by the Fonds voor Wetenschappelijk Onderzoek Vlaanderen (FWO) under contract ZKD1501-00-W01 and by the Belgian Science Policy Office under contract BR$/$143$/$A2$/$STARLAB.
The authors would like to thank Dr. Richard Stancliffe for the useful comments and interesting discussion that undoubtedly improved this manuscript.
A.E. is grateful to Glenn-Michael Oomen because he is always willing to discuss the evolution of binary systems.
L.S. is  senior FNRS research associate.
\end{acknowledgements}

\bibliographystyle{aa}
\bibliography{references}

\begin{thebibliography}{97}
\expandafter\ifx\csname natexlab\endcsname\relax\def\natexlab#1{#1}\fi

\bibitem[{{Abate} {et~al.}(2013){Abate}, {Pols}, {Izzard}, {Mohamed}, \& {de
  Mink}}]{Abate13}
{Abate}, C., {Pols}, O.~R., {Izzard}, R.~G., {Mohamed}, S.~S., \& {de Mink},
  S.~E. 2013, A\&A, 552, A26

\bibitem[{{Abate} {et~al.}(2018){Abate}, {Pols}, \& {Stancliffe}}]{Abate18}
{Abate}, C., {Pols}, O.~R., \& {Stancliffe}, R.~J. 2018, \aap, 620, A63

\bibitem[{{Aoki} {et~al.}(2008){Aoki}, {Beers}, {Sivarani}, {Marsteller},
  {Lee}, {Honda}, {Norris}, {Ryan}, \& {Carollo}}]{Aoki08}
{Aoki}, W., {Beers}, T.~C., {Sivarani}, T., {et~al.} 2008, \apj, 678, 1351

\bibitem[{{Artymowicz} {et~al.}(1991){Artymowicz}, {Clarke}, {Lubow}, \&
  {Pringle}}]{Artymowicz91}
{Artymowicz}, P., {Clarke}, C.~J., {Lubow}, S.~H., \& {Pringle}, J.~E. 1991,
  \apjl, 370, L35

\bibitem[{{Artymowicz} \& {Lubow}(1994)}]{ArtymowiczLubow94}
{Artymowicz}, P. \& {Lubow}, S.~H. 1994, ApJ, 421, 651

\bibitem[{{Asplund} {et~al.}(2009){Asplund}, {Grevesse}, {Sauval}, \&
  {Scott}}]{Asplund09}
{Asplund}, M., {Grevesse}, N., {Sauval}, A.~J., \& {Scott}, P. 2009, \araa, 47,
  481

\bibitem[{{Bidelman} \& {Keenan}(1951)}]{BidelmanKeenan51}
{Bidelman}, W.~P. \& {Keenan}, P.~C. 1951, ApJ, 114, 473

\bibitem[{{B{\"o}hm-Vitense} {et~al.}(2000){B{\"o}hm-Vitense}, {Carpenter},
  {Robinson}, {Ake}, \& {Brown}}]{Bohm-Vitense00}
{B{\"o}hm-Vitense}, E., {Carpenter}, K., {Robinson}, R., {Ake}, T., \& {Brown},
  J. 2000, \apj, 533, 969

\bibitem[{{Bona{\v c}i{\'c} Marinovi{\'c}} {et~al.}(2008){Bona{\v c}i{\'c}
  Marinovi{\'c}}, {Glebbeek}, \& {Pols}}]{BonacicMarinovic08}
{Bona{\v c}i{\'c} Marinovi{\'c}}, A.~A., {Glebbeek}, E., \& {Pols}, O.~R. 2008,
  A\&A, 480, 797

\bibitem[{{Bond}(1974)}]{Bond74}
{Bond}, H.~E. 1974, ApJ, 194, 95

\bibitem[{{Carquillat} {et~al.}(1998){Carquillat}, {Jorissen}, {Udry}, \&
  {Ginestet}}]{Carquillat98}
{Carquillat}, J.~M., {Jorissen}, A., {Udry}, S., \& {Ginestet}, N. 1998, \aaps,
  131, 49

\bibitem[{{Cseh} {et~al.}(2018){Cseh}, {Lugaro}, {D'Orazi}, {de Castro},
  {Pereira}, {Karakas}, {Moln{\'a}r}, {Plachy}, {Szab{\'o}}, {Pignatari}, \&
  {Cristallo}}]{Cseh18}
{Cseh}, B., {Lugaro}, M., {D'Orazi}, V., {et~al.} 2018, \aap, 620, A146

\bibitem[{{Davis} {et~al.}(2013){Davis}, {Siess}, \& {Deschamps}}]{Davis13}
{Davis}, P.~J., {Siess}, L., \& {Deschamps}, R. 2013, \aap, 556, A4

\bibitem[{{de Castro} {et~al.}(2016){de Castro}, {Pereira}, {Roig}, {Jilinski},
  {Drake}, {Chavero}, \& {Sales Silva}}]{deCastro16}
{de Castro}, D.~B., {Pereira}, C.~B., {Roig}, F., {et~al.} 2016, \mnras, 459,
  4299

\bibitem[{{de Ruyter} {et~al.}(2006){de Ruyter}, {van Winckel}, {Maas}, {Lloyd
  Evans}, {Waters}, \& {Dejonghe}}]{deRuyter06}
{de Ruyter}, S., {van Winckel}, H., {Maas}, T., {et~al.} 2006, \aap, 448, 641

\bibitem[{{Denissenkov} \& {Tout}(2000)}]{Denissenkov00}
{Denissenkov}, P.~A. \& {Tout}, C.~A. 2000, \mnras, 316, 395

\bibitem[{{Dermine} {et~al.}(2013){Dermine}, {Izzard}, {Jorissen}, \& {Van
  Winckel}}]{Dermine13}
{Dermine}, T., {Izzard}, R.~G., {Jorissen}, A., \& {Van Winckel}, H. 2013,
  A\&A, 551, A50

\bibitem[{{Escorza} {et~al.}(2017){Escorza}, {Boffin}, {Jorissen}, {Van Eck},
  {Siess}, {Van Winckel}, {Karinkuzhi}, {Shetye}, \& {Pourbaix}}]{Escorza17}
{Escorza}, A., {Boffin}, H.~M.~J., {Jorissen}, A., {et~al.} 2017, A\&A, 608,
  A100

\bibitem[{{Escorza} {et~al.}(2019{\natexlab{a}}){Escorza}, {Boffin},
  {Jorissen}, {Van Eck}, {Siess}, {Van Winckel}, {Karinkuzhi}, {Shetye}, \&
  {Pourbaix}}]{Escorza17err}
{Escorza}, A., {Boffin}, H.~M.~J., {Jorissen}, A., {et~al.} 2019{\natexlab{a}},
  \aap, 625, C3

\bibitem[{{Escorza} {et~al.}(2019{\natexlab{b}}){Escorza}, {Karinkuzhi},
  {Jorissen}, {Siess}, {Van Winckel}, {Pourbaix}, {Johnston}, {Miszalski},
  {Oomen}, {Abdul-Masih}, {Boffin}, {North}, {Manick}, {Shetye}, \&
  {Miko{\l}ajewska}}]{Escorza19}
{Escorza}, A., {Karinkuzhi}, D., {Jorissen}, A., {et~al.} 2019{\natexlab{b}},
  \aap, 626, A128

\bibitem[{{Ferguson} {et~al.}(2005){Ferguson}, {Alexander}, {Allard}, {Barman},
  {Bodnarik}, {Hauschildt}, {Heffner-Wong}, \& {Tamanai}}]{Ferguson05}
{Ferguson}, J.~W., {Alexander}, D.~R., {Allard}, F., {et~al.} 2005, \apj, 623,
  585

\bibitem[{{Gaia Collaboration} {et~al.}(2016{\natexlab{a}}){Gaia
  Collaboration}, {Brown}, {Vallenari}, {Prusti}, {de Bruijne}, {Mignard},
  {Drimmel}, {Babusiaux}, {Bailer-Jones}, {Bastian}, \&
  et~al.}]{Gaia-collaborationDR1}
{Gaia Collaboration}, {Brown}, A.~G.~A., {Vallenari}, A., {et~al.}
  2016{\natexlab{a}}, \aap, 595, A2

\bibitem[{{Gaia Collaboration} {et~al.}(2016{\natexlab{b}}){Gaia
  Collaboration}, {Prusti}, {de Bruijne}, {Brown}, {Vallenari}, {Babusiaux},
  {Bailer-Jones}, {Bastian}, {Biermann}, {Evans}, \& et~al.}]{GaiaMission}
{Gaia Collaboration}, {Prusti}, T., {de Bruijne}, J.~H.~J., {et~al.}
  2016{\natexlab{b}}, A\&A, 595, A1

\bibitem[{{Goriely} \& {Siess}(2018)}]{GorielySiess18}
{Goriely}, S. \& {Siess}, L. 2018, \aap, 609, A29

\bibitem[{{Gray} {et~al.}(2011){Gray}, {McGahee}, {Griffin}, \&
  {Corbally}}]{Gray11}
{Gray}, R.~O., {McGahee}, C.~E., {Griffin}, R.~E.~M., \& {Corbally}, C.~J.
  2011, AJ, 141, 160

\bibitem[{{Griffin} \& {Griffin}(1980)}]{Griffin80}
{Griffin}, R. \& {Griffin}, R. 1980, \mnras, 193, 957

\bibitem[{{Griffin}(1991)}]{Griffin91}
{Griffin}, R.~F. 1991, The Observatory, 111, 29

\bibitem[{{Griffin} {et~al.}(1996){Griffin}, {Jorissen}, \&
  {Mayor}}]{Griffin96}
{Griffin}, R.~F., {Jorissen}, A., \& {Mayor}, M. 1996, The Observatory, 116,
  298

\bibitem[{{Griffin} \& {Keenan}(1992)}]{Griffin92}
{Griffin}, R.~F. \& {Keenan}, P.~C. 1992, The Observatory, 112, 168

\bibitem[{{Han} {et~al.}(1995){Han}, {Eggleton}, {Podsiadlowski}, \&
  {Tout}}]{Han95}
{Han}, Z., {Eggleton}, P.~P., {Podsiadlowski}, P., \& {Tout}, C.~A. 1995,
  MNRAS, 277, 1443

\bibitem[{{Herwig} {et~al.}(1997){Herwig}, {Bloecker}, {Schoenberner}, \& {El
  Eid}}]{Herwig97}
{Herwig}, F., {Bloecker}, T., {Schoenberner}, D., \& {El Eid}, M. 1997, A\&A,
  324, L81

\bibitem[{{Hollands} {et~al.}(2018){Hollands}, {Tremblay}, {G{\"a}nsicke},
  {Gentile-Fusillo}, \& {Toonen}}]{Hollands18}
{Hollands}, M.~A., {Tremblay}, P.~E., {G{\"a}nsicke}, B.~T., {Gentile-Fusillo},
  N.~P., \& {Toonen}, S. 2018, \mnras, 480, 3942

\bibitem[{{Hurley} {et~al.}(2002){Hurley}, {Tout}, \& {Pols}}]{Hurley02}
{Hurley}, J.~R., {Tout}, C.~A., \& {Pols}, O.~R. 2002, \mnras, 329, 897

\bibitem[{{Husti} {et~al.}(2009){Husti}, {Gallino}, {Bisterzo}, {Straniero}, \&
  {Cristallo}}]{Husti09}
{Husti}, L., {Gallino}, R., {Bisterzo}, S., {Straniero}, O., \& {Cristallo}, S.
  2009, \pasa, 26, 176

\bibitem[{{Hut}(1981)}]{Hut81}
{Hut}, P. 1981, \aap, 99, 126

\bibitem[{{Iglesias} \& {Rogers}(1996)}]{IglesiasRogers96}
{Iglesias}, C.~A. \& {Rogers}, F.~J. 1996, \apj, 464, 943

\bibitem[{{Izzard} {et~al.}(2010){Izzard}, {Dermine}, \& {Church}}]{Izzard10}
{Izzard}, R.~G., {Dermine}, T., \& {Church}, R.~P. 2010, A\&A, 523, A10

\bibitem[{{Jeans}(1924)}]{Jeans24}
{Jeans}, J.~H. 1924, \mnras, 85, 2

\bibitem[{{Jeans}(1925)}]{Jeans25}
{Jeans}, J.~H. 1925, \mnras, 85, 912

\bibitem[{{Jorissen}(2004)}]{Jorissen04}
{Jorissen}, A. 2004, in Asymptotic Giant Branch Stars, Edited by Harm J. Habing
  and Hans Olofsson. Astronomy and Astrophysics Library. Springer, 2004.,
  461--518

\bibitem[{{Jorissen}(2020)}]{Jorissen19EWASS}
{Jorissen}, A. 2020, Mem. S.A.It, in press.

\bibitem[{{Jorissen} {et~al.}(2019){Jorissen}, {Boffin}, {Karinkuzhi}, {Van
  Eck}, {Escorza}, {Shetye}, \& {Van Winckel}}]{Jorissen19}
{Jorissen}, A., {Boffin}, H.~M.~J., {Karinkuzhi}, D., {et~al.} 2019, \aap, 626,
  A127

\bibitem[{{Jorissen} {et~al.}(1995){Jorissen}, {Hennen}, {Mayor}, {Bruch}, \&
  {Sterken}}]{Jorissen95}
{Jorissen}, A., {Hennen}, O., {Mayor}, M., {Bruch}, A., \& {Sterken}, C. 1995,
  \aap, 301, 707

\bibitem[{{Jorissen} \& {Mayor}(1988)}]{Jorissen88}
{Jorissen}, A. \& {Mayor}, M. 1988, A\&A, 198, 187

\bibitem[{{Jorissen} {et~al.}(1998){Jorissen}, {Van Eck}, {Mayor}, \&
  {Udry}}]{Jorissen98}
{Jorissen}, A., {Van Eck}, S., {Mayor}, M., \& {Udry}, S. 1998, A\&A, 332, 877

\bibitem[{{Jorissen} {et~al.}(2016){Jorissen}, {Van Eck}, {Van Winckel},
  {Merle}, {Boffin}, {Andersen}, {Nordstr{\"o}m}, {Udry}, {Masseron},
  {Lenaerts}, \& {Waelkens}}]{Jorissen16}
{Jorissen}, A., {Van Eck}, S., {Van Winckel}, H., {et~al.} 2016, A\&A, 586,
  A158

\bibitem[{{Kamath} {et~al.}(2016){Kamath}, {Wood}, {Van Winckel}, \&
  {Nie}}]{Kamath16}
{Kamath}, D., {Wood}, P.~R., {Van Winckel}, H., \& {Nie}, J.~D. 2016, \aap,
  586, L5

\bibitem[{{Karinkuzhi} {et~al.}(2018){Karinkuzhi}, {Van Eck}, {Jorissen},
  {Goriely}, {Siess}, {Merle}, {Escorza}, {Van der Swaelmen}, {Boffin},
  {Masseron}, {Shetye}, \& {Plez}}]{Karinkuzhi18}
{Karinkuzhi}, D., {Van Eck}, S., {Jorissen}, A., {et~al.} 2018, \aap, 618, A32

\bibitem[{{Keenan}(1942)}]{Keenan42}
{Keenan}, P.~C. 1942, ApJ, 96, 101

\bibitem[{{Kilic} {et~al.}(2018){Kilic}, {Hambly}, {Bergeron},
  {Genest-Beaulieu}, \& {Rowell}}]{Kilic18}
{Kilic}, M., {Hambly}, N.~C., {Bergeron}, P., {Genest-Beaulieu}, C., \&
  {Rowell}, N. 2018, \mnras, 479, L113

\bibitem[{{Kleinman} {et~al.}(2013){Kleinman}, {Kepler}, {Koester}, {Pelisoli},
  {Pe{\c c}anha}, {Nitta}, {Costa}, {Krzesinski}, {Dufour}, {Lachapelle},
  {Bergeron}, {Yip}, {Harris}, {Eisenstein}, {Althaus}, \&
  {C{\'o}rsico}}]{Kleinman13}
{Kleinman}, S.~J., {Kepler}, S.~O., {Koester}, D., {et~al.} 2013, \apjs, 204, 5

\bibitem[{{Lindegren} {et~al.}(2018){Lindegren}, {Hern{\'a}ndez}, {Bombrun},
  {Klioner}, {Bastian}, {Ramos-Lerate}, {de Torres}, {Steidelm{\"u}ller},
  {Stephenson}, {Hobbs}, {Lammers}, {Biermann}, {Geyer}, {Hilger}, {Michalik},
  {Stampa}, {McMillan}, {Casta{\~n}eda}, {Clotet}, {Comoretto}, {Davidson},
  {Fabricius}, {Gracia}, {Hambly}, {Hutton}, {Mora}, {Portell}, {van Leeuwen},
  {Abbas}, {Abreu}, {Altmann}, {Andrei}, {Anglada}, {Balaguer-N{\'u}{\~n}ez},
  {Barache}, {Becciani}, {Bertone}, {Bianchi}, {Bouquillon}, {Bourda},
  {Br{\"u}semeister}, {Bucciarelli}, {Busonero}, {Buzzi}, {Cancelliere},
  {Carlucci}, {Charlot}, {Cheek}, {Crosta}, {Crowley}, {de Bruijne}, {de
  Felice}, {Drimmel}, {Esquej}, {Fienga}, {Fraile}, {Gai}, {Garralda},
  {Gonz{\'a}lez-Vidal}, {Guerra}, {Hauser}, {Hofmann}, {Holl}, {Jordan},
  {Lattanzi}, {Lenhardt}, {Liao}, {Licata}, {Lister}, {L{\"o}ffler},
  {Marchant}, {Martin-Fleitas}, {Messineo}, {Mignard}, {Morbidelli}, {Poggio},
  {Riva}, {Rowell}, {Salguero}, {Sarasso}, {Sciacca}, {Siddiqui}, {Smart},
  {Spagna}, {Steele}, {Taris}, {Torra}, {van Elteren}, {van Reeven}, \&
  {Vecchiato}}]{Lindegren18}
{Lindegren}, L., {Hern{\'a}ndez}, J., {Bombrun}, A., {et~al.} 2018, \aap, 616,
  A2

\bibitem[{{Lindegren} {et~al.}(2016){Lindegren}, {Lammers}, {Bastian},
  {Hern{\'a}ndez}, {Klioner}, {Hobbs}, {Bombrun}, {Michalik}, {Ramos-Lerate},
  {Butkevich}, {Comoretto}, {Joliet}, {Holl}, {Hutton}, {Parsons},
  {Steidelm{\"u}ller}, {Abbas}, {Altmann}, {Andrei}, {Anton}, {Bach},
  {Barache}, {Becciani}, {Berthier}, {Bianchi}, {Biermann}, {Bouquillon},
  {Bourda}, {Br{\"u}semeister}, {Bucciarelli}, {Busonero}, {Carlucci},
  {Casta{\~n}eda}, {Charlot}, {Clotet}, {Crosta}, {Davidson}, {de Felice},
  {Drimmel}, {Fabricius}, {Fienga}, {Figueras}, {Fraile}, {Gai}, {Garralda},
  {Geyer}, {Gonz{\'a}lez-Vidal}, {Guerra}, {Hambly}, {Hauser}, {Jordan},
  {Lattanzi}, {Lenhardt}, {Liao}, {L{\"o}ffler}, {McMillan}, {Mignard}, {Mora},
  {Morbidelli}, {Portell}, {Riva}, {Sarasso}, {Serraller}, {Siddiqui}, {Smart},
  {Spagna}, {Stampa}, {Steele}, {Taris}, {Torra}, {van Reeven}, {Vecchiato},
  {Zschocke}, {de Bruijne}, {Gracia}, {Raison}, {Lister}, {Marchant},
  {Messineo}, {Soffel}, {Osorio}, {de Torres}, \& {O'Mullane}}]{Lindegren16}
{Lindegren}, L., {Lammers}, U., {Bastian}, U., {et~al.} 2016, \aap, 595, A4

\bibitem[{{Livio} \& {Soker}(1988)}]{LivioSoker88}
{Livio}, M. \& {Soker}, N. 1988, \apj, 329, 764

\bibitem[{{Lugaro} {et~al.}(2016){Lugaro}, {Campbell}, {D'Orazi}, {Karakas},
  {Garcia-Hernandez}, {Stancliffe}, {Tagliente}, {Iliadis}, \&
  {Rauscher}}]{Lugaro16}
{Lugaro}, M., {Campbell}, S.~W., {D'Orazi}, V., {et~al.} 2016, JPCS, 665,
  012021

\bibitem[{{Matrozis} \& {Stancliffe}(2016)}]{Matrozis16}
{Matrozis}, E. \& {Stancliffe}, R.~J. 2016, \aap, 592, A29

\bibitem[{{Matrozis} \& {Stancliffe}(2017)}]{Matrozis17}
{Matrozis}, E. \& {Stancliffe}, R.~J. 2017, \aap, 606, A55

\bibitem[{{McClure} \& {Woodsworth}(1990)}]{McClureWoodsworth90}
{McClure}, R.~D. \& {Woodsworth}, A.~W. 1990, ApJ, 352, 709

\bibitem[{{Mermilliod} {et~al.}(2007){Mermilliod}, {Andersen}, {Latham}, \&
  {Mayor}}]{Mermilliod07}
{Mermilliod}, J.-C., {Andersen}, J., {Latham}, D.~W., \& {Mayor}, M. 2007,
  \aap, 473, 829

\bibitem[{{Nie} {et~al.}(2017){Nie}, {Wood}, \& {Nicholls}}]{Nie17}
{Nie}, J.~D., {Wood}, P.~R., \& {Nicholls}, C.~P. 2017, \apj, 835, 209

\bibitem[{{North} {et~al.}(2000){North}, {Jorissen}, \& {Mayor}}]{North00}
{North}, P., {Jorissen}, A., \& {Mayor}, M. 2000, in IAU Symposium, Vol. 177,
  The Carbon Star Phenomenon, ed. R.~F. {Wing (Dordrecht: Kluwer)}, 269

\bibitem[{{North} {et~al.}(2020){North}, {Jorissen}, {Escorza}, {Miszalski}, \&
  {Mikolajewska}}]{North20}
{North}, P.~L., {Jorissen}, A., {Escorza}, A., {Miszalski}, B., \&
  {Mikolajewska}, J. 2020, arXiv e-prints, arXiv:2001.11319

\bibitem[{{Oomen} {et~al.}(2020){Oomen}, {Pols}, {Van Winckel}, \&
  {Nelemans}}]{Oomen20}
{Oomen}, G.-M., {Pols}, O., {Van Winckel}, H., \& {Nelemans}, G. 2020, \aap, in
  prep.

\bibitem[{{Oomen} {et~al.}(2018){Oomen}, {Van Winckel}, {Pols}, {Nelemans},
  {Escorza}, {Manick}, {Kamath}, \& {Waelkens}}]{Oomen18}
{Oomen}, G.-M., {Van Winckel}, H., {Pols}, O., {et~al.} 2018, \aap, 620, A85

\bibitem[{{Paczynski}(1976)}]{Paczynski76}
{Paczynski}, B. 1976, in IAU Symposium, Vol.~73, Structure and Evolution of
  Close Binary Systems, ed. P.~{Eggleton}, S.~{Mitton}, \& J.~{Whelan}, 75

\bibitem[{{Pols} {et~al.}(2003){Pols}, {Karakas}, {Lattanzio}, \&
  {Tout}}]{Pols03}
{Pols}, O.~R., {Karakas}, A.~I., {Lattanzio}, J.~C., \& {Tout}, C.~A. 2003, in
  Astronomical Society of the Pacific Conference Series, Vol. 303, Symbiotic
  Stars Probing Stellar Evolution, ed. R.~L.~M. {Corradi}, J.~{Mikolajewska},
  \& T.~J. {Mahoney}, 290

\bibitem[{{Pourbaix} \& {Jorissen}(2000)}]{PJ2000}
{Pourbaix}, D. \& {Jorissen}, A. 2000, A\&AS, 145, 161

\bibitem[{{Proffitt} \& {Michaud}(1989)}]{Proffitt89}
{Proffitt}, C.~R. \& {Michaud}, G. 1989, \apj, 345, 998

\bibitem[{{Saladino} \& {Pols}(2019)}]{Saladino19}
{Saladino}, M.~I. \& {Pols}, O.~R. 2019, \aap, 629, A103

\bibitem[{{Saladino} {et~al.}(2018){Saladino}, {Pols}, {van der Helm},
  {Pelupessy}, \& {Portegies Zwart}}]{Saladino18}
{Saladino}, M.~I., {Pols}, O.~R., {van der Helm}, E., {Pelupessy}, I., \&
  {Portegies Zwart}, S. 2018, \aap, 618, A50

\bibitem[{{Schr{\"o}der} \& {Cuntz}(2005)}]{SchroderCuntz05}
{Schr{\"o}der}, K.~P. \& {Cuntz}, M. 2005, \apjl, 630, L73

\bibitem[{{Schr{\"o}der} \& {Cuntz}(2007)}]{SchroderCuntz07}
{Schr{\"o}der}, K.-P. \& {Cuntz}, M. 2007, A\&A, 465, 593

\bibitem[{{Shen}(2015)}]{Shen15}
{Shen}, K.~J. 2015, \apjl, 805, L6

\bibitem[{{Shen} \& {Moore}(2014)}]{Shen14}
{Shen}, K.~J. \& {Moore}, K. 2014, \apj, 797, 46

\bibitem[{{Shetye} {et~al.}(2020){Shetye}, {Van Eck}, {Goriely}, {Siess},
  {Jorissen}, {Escorza}, \& {Van Winckel}}]{Shetye20}
{Shetye}, S., {Van Eck}, S., {Goriely}, S., {et~al.} 2020, \aap, 635, L6

\bibitem[{{Shetye} {et~al.}(2018){Shetye}, {Van Eck}, {Jorissen}, {Van
  Winckel}, {Siess}, {Goriely}, {Escorza}, {Karinkuzhi}, \& {Plez}}]{Shetye18}
{Shetye}, S., {Van Eck}, S., {Jorissen}, A., {et~al.} 2018, \aap, 620, A148

\bibitem[{{Siess}(2006)}]{Siess2006}
{Siess}, L. 2006, \aap, 448, 717

\bibitem[{{Siess} {et~al.}(2014){Siess}, {Davis}, \& {Jorissen}}]{Siess14}
{Siess}, L., {Davis}, P.~J., \& {Jorissen}, A. 2014, \aap, 565, A57

\bibitem[{{Siess} {et~al.}(2013){Siess}, {Izzard}, {Davis}, \&
  {Deschamps}}]{Siess13}
{Siess}, L., {Izzard}, R.~G., {Davis}, P.~J., \& {Deschamps}, R. 2013, \aap,
  550, A100

\bibitem[{{Smith} \& {Lambert}(1988)}]{SmithLambert88}
{Smith}, V.~V. \& {Lambert}, D.~L. 1988, \apj, 333, 219

\bibitem[{{Smith} \& {Lambert}(1990)}]{SmithLambert90}
{Smith}, V.~V. \& {Lambert}, D.~L. 1990, \apjs, 72, 387

\bibitem[{{Soker}(2000)}]{Soker00}
{Soker}, N. 2000, A\&A, 357, 557

\bibitem[{{Stancliffe} {et~al.}(2009){Stancliffe}, {Church}, {Angelou}, \&
  {Lattanzio}}]{Stancliffe09}
{Stancliffe}, R.~J., {Church}, R.~P., {Angelou}, G.~C., \& {Lattanzio}, J.~C.
  2009, \mnras, 396, 2313

\bibitem[{{Stancliffe} \& {Glebbeek}(2008)}]{Stancliffe08}
{Stancliffe}, R.~J. \& {Glebbeek}, E. 2008, \mnras, 389, 1828

\bibitem[{{Stancliffe} {et~al.}(2007){Stancliffe}, {Glebbeek}, {Izzard}, \&
  {Pols}}]{Stancliffe07}
{Stancliffe}, R.~J., {Glebbeek}, E., {Izzard}, R.~G., \& {Pols}, O.~R. 2007,
  \aap, 464, L57

\bibitem[{{Tout} \& {Eggleton}(1988)}]{ToutEggleton88}
{Tout}, C.~A. \& {Eggleton}, P.~P. 1988, MNRAS, 231, 823

\bibitem[{{Townsley} {et~al.}(2019){Townsley}, {Miles}, {Shen}, \&
  {Kasen}}]{Townsley19}
{Townsley}, D.~M., {Miles}, B.~J., {Shen}, K.~J., \& {Kasen}, D. 2019, \apjl,
  878, L38

\bibitem[{{Tremblay} {et~al.}(2019){Tremblay}, {Cukanovaite}, {Gentile
  Fusillo}, {Cunningham}, \& {Hollands}}]{Tremblay19}
{Tremblay}, P.~E., {Cukanovaite}, E., {Gentile Fusillo}, N.~P., {Cunningham},
  T., \& {Hollands}, M.~A. 2019, \mnras, 482, 5222

\bibitem[{{Udry} {et~al.}(1998{\natexlab{a}}){Udry}, {Jorissen}, {Mayor}, \&
  {Van Eck}}]{Udry98}
{Udry}, S., {Jorissen}, A., {Mayor}, M., \& {Van Eck}, S. 1998{\natexlab{a}},
  A\&AS, 131, 25

\bibitem[{{Udry} {et~al.}(1998{\natexlab{b}}){Udry}, {Mayor}, {Van Eck},
  {Jorissen}, {Prevot}, {Grenier}, \& {Lindgren}}]{Udry98II}
{Udry}, S., {Mayor}, M., {Van Eck}, S., {et~al.} 1998{\natexlab{b}}, A\&AS,
  131, 43

\bibitem[{{Van der Swaelmen} {et~al.}(2017){Van der Swaelmen}, {Boffin},
  {Jorissen}, \& {Van Eck}}]{VanderSwaelmen17}
{Van der Swaelmen}, M., {Boffin}, H.~M.~J., {Jorissen}, A., \& {Van Eck}, S.
  2017, \aap, 597, A68

\bibitem[{{Van Winckel}(2017)}]{vanWinckel17}
{Van Winckel}, H. 2017, in IAU Symposium, Vol. 323, Planetary Nebulae:
  Multi-Wavelength Probes of Stellar and Galactic Evolution, ed. X.~{Liu},
  L.~{Stanghellini}, \& A.~{Karakas}, 231--234

\bibitem[{{Vos} {et~al.}(2015){Vos}, {{\O}stensen}, {Marchant}, \& {Van
  Winckel}}]{Vos15}
{Vos}, J., {{\O}stensen}, R.~H., {Marchant}, P., \& {Van Winckel}, H. 2015,
  A\&A, 579, A49

\bibitem[{{Warner}(1965)}]{Warner1965}
{Warner}, B. 1965, \mnras, 129, 263

\bibitem[{{Wolf} {et~al.}(2013){Wolf}, {Bildsten}, {Brooks}, \&
  {Paxton}}]{Wolf13}
{Wolf}, W.~M., {Bildsten}, L., {Brooks}, J., \& {Paxton}, B. 2013, \apj, 777,
  136

\bibitem[{{Zahn}(1977)}]{Zahn77}
{Zahn}, J.~P. 1977, \aap, 500, 121

\bibitem[{{Zahn}(1989)}]{Zahn89}
{Zahn}, J.-P. 1989, \aap, 220, 112

\end{thebibliography}

\clearpage
\onecolumn
\begin{appendix}

\section{Periods and eccentricities of the comparison sample}\label{sec:app}

Table \ref{orbits} collects periods, eccentricities, and masses of the Ba, CH and S giant systems included in Fig. \ref{elogp_giants}. We list the original references to the orbital elements in the table, while the masses come from \cite{Jorissen19}, \cite{Shetye18}, or have been computed by us following their methods. We do not include the main-sequence systems because they all come from \cite{Escorza19} and \cite{North20}.

\LTcapwidth=0.75\textwidth
\renewcommand\arraystretch{1.5}
\begin{longtable}{lccccc}
\caption{\label{orbits} Period, eccentricity, original references to these and masses of the giant systems displayed in Fig. \ref{elogp_giants}. \\
\textbf{References:} (0) Fit improved for this work; (1) \cite{Udry98}; (2) \cite{Jorissen95}; (3) \cite{Jorissen19}; (4) \cite{Griffin96}; (5) \cite{Udry98II}; (6) \cite{McClureWoodsworth90}; (7) \cite{VanderSwaelmen17}; (8) \cite{Griffin80}; (9) \cite{Griffin91}; (10) \cite{Griffin92}; (11) \cite{Jorissen98}; (12) \cite{Jorissen98}; (13) \cite{Mermilliod07}; (14) \cite{Jorissen16}; (15) \cite{Carquillat98}.}\\
\hline\hline
\rule[0mm]{0mm}{4mm}
Name & type & $P$ [days] & $e$ & Orbit Ref. & Mass [M$_{\odot}$]\\
\hline
\endfirsthead
\hline\hline
\rule[0mm]{0mm}{4mm}
Name & type & $P$ [days] & $e$ & Orbit Ref. & Mass [M$_{\odot}$]\\
\hline
\endhead
DM-64$^{\circ}$4333 & gBa & 386.04 $\pm$ 0.5 & 0.03 $\pm$ 0.01 & 1 & 1.4$^{+0.1}_{-0.1}$ \\
DM-42$^{\circ}$2048 & gBa & 3260 $\pm$ 28.3 & 0.08 $\pm$ 0.02 & 1 & 1.9$^{+0.7}_{-0.5}$ \\
DM-14$^{\circ}$2678 & gBa & 3470.5 $\pm$ 107.7 & 0.22 $\pm$ 0.04 & 1 & 3.0$^{+0.2}_{-0.2}$ \\
DM-01$^{\circ}$3022 & gBa & 3252.53 $\pm$ 31.4 & 0.28 $\pm$ 0.02 & 1 & 1.6$^{+0.1}_{-0.1}$ \\
HD\,5424 & gBa & 1881.5 $\pm$ 18.6 & 0.23 $\pm$ 0.04 & 1 & 1.3$^{+0.4}_{-0.3}$ \\
HD\,16458 & gBa & 2018 $\pm$ 12 & 0.1 $\pm$ 0.02 & 2 & 1.9$^{+0.1}_{-0.1}$ \\
HD\,18182 & gBa & 8059 $\pm$ 256 & 0.31 $\pm$ 0.13 & 3,0 & 1.8$^{+0.2}_{-0.1}$ \\
HD\,20394 & gBa & 2226 $\pm$ 22 & 0.2 $\pm$ 0.03 & 4 & 2.0$^{+0.2}_{-0.2}$ \\
HD\,24035 & gBa & 377.8 $\pm$ 0.3 & 0.02 $\pm$ 0.01 & 1 & 1.3$^{+0.3}_{-0.2}$ \\
HD\,27271 & gBa & 1693.8 $\pm$ 9.1 & 0.22 $\pm$ 0.02 & 5 & 2.9$^{+0.2}_{-0.2}$ \\
HD\,31487 & gBa & 1066.4 $\pm$ 2.6 & 0.05 $\pm$ 0.01 & 6 & 3.4$^{+0.2}_{-0.3}$ \\
HD\,40430 & gBa & 5609 $\pm$ 55 & 0.22 $\pm$ 0.01 & 3 & 2.3$^{+0.2}_{-0.2}$ \\
HD\,43389 & gBa & 1689 $\pm$ 8.7 & 0.08 $\pm$ 0.02 & 5 & 1.8$^{+0.4}_{-0.3}$ \\
HD\,44896 & gBa & 628.89 $\pm$ 0.9 & 0.02 $\pm$ 0.01 & 5 & 3.0$^{+1.2}_{-1.0}$ \\
HD\,46407 & gBa & 457.43 $\pm$ 0.10 & 0.013 $\pm$ 0.008 & 5 & 2.1$^{+0.6}_{-0.7}$ \\
HD\,49641 & gBa & 1785 $\pm$ 54 & 0.07 $\pm$ 0.1 & 6,7 & 2.7$^{+1.2}_{-0.8}$ \\
HD\,49841 & gBa & 897.1 $\pm$ 1.8 & 0.16 $\pm$ 0.01 & 5 & 2.8$^{+0.2}_{-0.2}$ \\
HD\,50082 & gBa & 2896 $\pm$ 21.3 & 0.19 $\pm$ 0.02 & 5 & 1.6$^{+0.3}_{-0.2}$ \\
HD\,51959 & gBa & 9717 $\pm$ 157 & 0.53 $\pm$ 0.04 & 3 & 1.2$^{+0.1}_{-0.1}$ \\
HD\,53199 & gBa & 8314 $\pm$ 99 & 0.24 $\pm$ 0.01 & 3 & 2.5$^{+0.1}_{-0.1}$ \\
HD\,58121 & gBa & 1214.3 $\pm$ 5.7 & 0.14 $\pm$ 0.02 & 5 & 2.6$^{+0.5}_{-0.4}$ \\
HD\,58368 & gBa & 672.7 $\pm$ 1.3 & 0.22 $\pm$ 0.02 & 6 & 2.6$^{+0.1}_{-0.2}$ \\
HD\,59852 & gBa & 3463.91 $\pm$ 53.8 & 0.15 $\pm$ 0.06 & 1 & 2.5$^{+0.2}_{-0.3}$ \\
HD\,77247 & gBa & 80.53 $\pm$ 0.01 & 0.09 $\pm$ 0.01 & 6 & 3.9$^{+0.1}_{-0.2}$ \\
HD\,84678 & gBa & 1629.91 $\pm$ 10.4 & 0.06 $\pm$ 0.02 & 1 & 2.3$^{+0.6}_{-0.5}$ \\
HD\,88562 & gBa & 1445 $\pm$ 8.5 & 0.2 $\pm$ 0.02 & 1 & 1.0$^{+0.1}_{-0.1}$ \\
HD\,91208 & gBa & 1754 $\pm$ 13.3 & 0.17 $\pm$ 0.02 & 1 & 2.3$^{+0.1}_{-0.2}$ \\
HD\,92626 & gBa & 918.2 $\pm$ 1.2 & 0 $\pm$ 0.01 & 5 & 3.1$^{+0.4}_{-0.6}$ \\
HD\,95193 & gBa & 1653.7 $\pm$ 9 & 0.13 $\pm$ 0.02 & 1 & 2.7$^{+0.1}_{-0.1}$ \\
HD\,98839 & gBa & 16471 $\pm$ 113 & 0.56 $\pm$ 0.005 & 3 & 4.3$^{+0.2}_{-0.2}$ \\
HD\,101013 & gBa & 1711 $\pm$ 4 & 0.2 $\pm$ 0.01 & 6,8 & 1.7$^{+0.3}_{-0.3}$ \\
HD\,104979 & gBa & 19295 $\pm$ - & 0.08 $\pm$ - & 3,0 & 2.7$^{+0.1}_{-0.2}$ \\
HD\,107541 & gBa & 3569.92 $\pm$ 46.1 & 0.1 $\pm$ 0.03 & 5 & 1.1$^{+0.2}_{-0.1}$ \\
HD\,119185 & gBa & 22065 $\pm$ - & 0.6 $\pm$ - & 3,0 & 1.7$^{+0.2}_{-0.2}$ \\
HD\,121447 & gBa & 185.7 $\pm$ 0.1 & 0.015 $\pm$ 0.013 & 2 & 1.6$^{+0.1}_{-0.1}$ \\
HD\,123949 & gBa & 8523 $\pm$ 8 & 0.9162 $\pm$ 0.0003 & 3 & 1.3$^{+0.3}_{-0.1}$ \\
HD\,134698 & gBa & 10005 $\pm$ - & 0.95 $\pm$ - & 3,0 & 1.5$^{+0.2}_{-0.2}$ \\
HD\,139195 & gBa & 5324 $\pm$ 19 & 0.35 $\pm$ 0.02 & 9 & 2.6$^{+0.1}_{-0.1}$ \\
HD\,143899 & gBa & 1461.6 $\pm$ 6.9 & 0.19 $\pm$ 0.02 & 1 & 2.4$^{+0.1}_{-0.1}$ \\
HD\,154430 & gBa & 1668 $\pm$ 17 & 0.11 $\pm$ 0.03 & 1 & 2.3$^{+1.4}_{-0.7}$ \\
HD\,178717 & gBa & 2866 $\pm$ 21 & 0.43 $\pm$ 0.03 & 6 & 1.6$^{+0.9}_{-0.7}$ \\
HD\,180622 & gBa & 4049 $\pm$ 37.7 & 0.06 $\pm$ 0.1 & 7 & 1.8$^{+0.3}_{-0.2}$ \\
HD\,183915 & gBa & 4382 $\pm$ 21 & 0.27 $\pm$ 0.02 & 3 & 1.8$^{+1.0}_{-0.6}$ \\
HD\,196673 & gBa & 7780 $\pm$ 117 & 0.59 $\pm$ 0.02 & 3 & 5.0$^{+0.0}_{-0.1}$ \\
HD\,199939 & gBa & 584.9 $\pm$ 0.7 & 0.28 $\pm$ 0.01 & 6 & 2.8$^{+0.4}_{-0.4}$ \\
HD\,200063 & gBa & 1735.45 $\pm$ 8.16 & 0.07 $\pm$ 0.04 & 5 & 2.0$^{+1.3}_{-0.9}$ \\
HD\,201657 & gBa & 1710.4 $\pm$ 15 & 0.17 $\pm$ 0.07 & 5 & 1.8$^{+0.5}_{-0.4}$ \\
HD\,201824 & gBa & 2837 $\pm$ 13 & 0.34 $\pm$ 0.02 & 4 & 1.7$^{+0.4}_{-0.2}$ \\
HD\,202109 & gBa & 6489 $\pm$ 31 & 0.22 $\pm$ 0.03 & 10 & 3.4$^{+0.2}_{-0.4}$ \\
HD\,204075 & gBa & 2378.23 $\pm$ 55.6 & 0.28 $\pm$ 0.07 & 6,11 & 4.5$^{+0.3}_{-0.2}$ \\
HD\,205011 & gBa & 2837 $\pm$ 10 & 0.24 $\pm$ 0.02 & 6,11 & 1.8$^{+0.3}_{-0.3}$ \\
HD\,210946 & gBa & 1529.5 $\pm$ 4.1 & 0.13 $\pm$ 0.01 & 5 & 1.8$^{+0.5}_{-0.4}$ \\
HD\,211594 & gBa & 1018.9 $\pm$ 2.7 & 0.06 $\pm$ 0.01 & 5 & 2.0$^{+0.3}_{-0.2}$ \\
HD\,218356 & gBa & 111.155 $\pm$ 0.017 & 0.0 $\pm$ 0.0 & 12 & 4.3$^{+0.2}_{-1.1}$ \\
HD\,223617 & gBa & 1293.7 $\pm$ 3.9 & 0.06 $\pm$ 0.02 & 5,6 & 1.4$^{+0.1}_{-0.1}$ \\
NGC\,2420-173 & gBa & 1479 $\pm$ 9.1 & 0.43 $\pm$ 0.05 & 13 & 3.0$^{+0.3}_{-0.4}$ \\
NGC\,2420-250 & gBa & 1403.6 $\pm$ 3.5 & 0.08 $\pm$ 0.03 & 13 & 2.096$^{+0.005}_{-0.005}$ \\
\hline
BD+41$^{\circ}$2150 & gCH & 322.84 $\pm$ 0.08 & 0.055 $\pm$ 0.001 & 14 & 1.27$^{+0.2}_{-0.2}$ \\
HD\,26 & gCH & 19634 $\pm$ 2812 & 0.08 $\pm$ 0.03 & 14 & 2.2$^{+0.2}_{-0.4}$ \\
HD\,5223 & gCH & 755.2 $\pm$ 3.4 & 0 & 6 & 1.4$^{+0.5}_{-0.4}$ \\
HD\,198269 & gCH & 1295 $\pm$ 10 & 0.094 $\pm$ 0.018 & 6 & 4.17$^{+0.08}_{-0.29}$ \\
HD\,201626 & gCH & 1465 $\pm$ 15 & 0.103 $\pm$ 0.038 & 6 & 1.9$^{+0.3}_{-0.3}$ \\
HD\,209621 & gCH & 407.4 $\pm$ 1.1 & 0 & 6 & 3.6$^{+0.6}_{-1.5}$ \\
HD\,224959 & gCH & 1273 $\pm$ 14 & 0.179 $\pm$ 0.038 & 6 & 2.4$^{+0.5}_{-0.5}$ \\
\hline
BD+28$^{\circ}$4592 & S & 1252.9 $\pm$ 3.5 & 0.09 $\pm$ 0.02 & 1 & 2 \\
HD\,63733 & S & 1160.7 $\pm$ 8.9 & 0.23 $\pm$ 0.03 & 1 & 2.5 \\
HD\,170970 & S & 4651 $\pm$ 10 & 0.19 $\pm$ 0.01 & 3 & 1.5 \\
HD\,189581 & S & 618 $\pm$ 1 & 0.0 $\pm$ 0.2 & 3 & 2 \\
HD\,191226 & S & 1210.4 $\pm$ 4.3 & 0.19 $\pm$ 0.02 & 15 & 3.5 \\
HD\,191589 & S & 377.3 $\pm$ 0.1 & 0.250 $\pm$ 0.003 & 1 & 1 \\
HD\,215336 & S & 1143.6 $\pm$ 0.7 & 0.040 $\pm$ 0.009 & 3 & 1.5 \\
\hline\hline
\end{longtable}

\end{appendix}

\end{document}